\documentclass[twocolumn,preprint]{aastex631}

\usepackage{array}
\usepackage{tabularx}
\usepackage{ulem}

\begin{document}

\title{An Afterglow Study of the ``New Year's Burst" GRB 220101A}

\author[0000-0003-1101-8436]{Agniva Roychowdhury}
\affiliation{National Centre for Radio Astrophysics - Tata Institute of Fundamental Research, Ganeshkhind, Pune 411007, MH, India}
\affiliation{Space Telescope Science Institute, 3700 San Martin Drive, Baltimore, MD 21218, USA}
\email{agniva.physics@gmail.com}

\author[0000-0002-5477-0217]{Tuomas Kangas}
\affiliation{Finnish Centre for Astronomy with ESO (FINCA), FI-20014 University of Turku, Finland}
\affiliation{Department of Physics and Astronomy, FI-20014 University of Turku, Finland}

\author[0000-0002-6652-9279]{Andrew Fruchter}
\affiliation{Space Telescope Science Institute, 3700 San Martin Drive, Baltimore, MD 21218, USA}

\author[0000-0000-0000-0000]{A. Pe'er}
\affiliation{Department of Physics, Bar Ilan University, Ramat-Gan 52900, Israel}

\author[0000-0000-0000-0000]{K. Bhirombhakdi}
\affiliation{Space Telescope Science Institute, 3700 San Martin Drive, Baltimore, MD 21218, USA}
\affiliation{Lead Prompt Security Researcher $|$ AI Safety $\&$ Red-Teaming at InjectPrompt}

\author[0000-0000-0000-0000]{J. Graham}
\affiliation{Visitor, University of Hawaii at Manoa, 2505 Correa Rd, Honolulu, HI 96822}

\author[0000-0003-1637-267X]{K. Misra}
\affiliation{Aryabhatta Research Institute of Observational Sciences, Manora Peak, Nainital 263001, India}

\author[0000-0001-7821-9369]{A.J Levan}
\affiliation{Department of Astrophysics/IMAPP, Radboud University Ni\nobreak{}jmegen, P.O.~Box 9010, Nijmegen, 6500~GL, The Netherlands}

\author[0000-0000-0000-0000]{B. Cenko}
\affiliation{NASA Goddard Space Flight Center, Greenbelt, MD 20771, USA}

\author[0000-0001-6455-5660]{A. Cucchiara}
\affiliation{NASA Headquarters, 300 E St. SW, Washington DC USA}

\author[0000-0000-0000-0000]{V. Cunningham}
\affiliation{University of Maryland, College Park, MD 20742, USA}

\author[0000-0002-5826-0548]{B. P. Gompertz}
\affiliation{School of Physics and Astronomy, University of Birmingham, Birmingham B15 2TT, United Kingdom}

\affiliation{Institute for Gravitational Wave Astronomy, University of Birmingham, Birmingham B15 2TT, United Kingdom}

%\author[0000-0000-0000-0000]{L. Resmi}
%\affiliation{Indian Institute of Space Science and Technology India, J2GM+HGJ, Valiamala Road, Valiamala, Kerala 695541, India}

\author[0000-0000-0000-0000]{D. Perley}
\affiliation{Liverpool John Moores University (LJMU), 10 Copperas Hl, Liverpool L3 5AH, United Kingdom}

\author[0000-0000-0000-0000]{J. Racusin}
\affiliation{NASA Goddard Space Flight Center, Greenbelt, MD 20771, USA}

\author[0000-0003-3274-6336]{N. R. Tanvir}
\affiliation{School of Physics and Astronomy, University of Leicester, University Rd, Leicester LE1 7RH, United Kingdom}

\begin{abstract}

We present a detailed broadband afterglow study of GRB 220101A ($10^4\lesssim\Delta T\lesssim10^7$ s) combining multi-wavelength data from soft X-rays until 6 GHz. The afterglow light curves in both X-ray and optical show distinct steepening around $\sim9$ days, followed by a sharp post-break decay index of $\sim2.99\pm0.10$. We fit the light curves using the afterglow modelling package \texttt{afterglowpy} for both Top-hat and Gaussian jets for different values of the electronic participation fraction $\xi$ from 0.01 to 1.0 and find that, although the radio behavior is well described by the $\xi=1.0$ case, the required circumburst medium (CBM) densities are very low, $<10^{-4}$ cm$^{-3}$. However, the resulting energy requirements are modest, $\sim10^{52}$ erg, with an electron energy distribution (EED) index $p\sim2.05$. Similar results are also obtained from an analytic model fit to the light curve, except the predicted $p$ is higher, $\sim2.40$. The observed post-break decay index of $2.99$ is at least 5$\sigma$ away from $p$, which is one of the steepest observed so far. We also find that when ignoring the radio observations, the CBM density is raised by a few orders of magnitude to $\sim0.01$ cm$^{-3}$ for $\xi=1.0$, still far from the expected ISM density ($>1$ cm$^{-3}$) of GRB environments, which are highly star forming regions. Similarly low ISM densities have been seen in modeling of other LAT GRBs as well, especially ones with reverse-shock features (e.g., GRBs 130427A, 160509A and 160625B), thereby hinting at either an issue with the standard model or possible evacuated cavities where GRBs explode.

%\textcolor{blue}{mention narrow jet?}

\end{abstract}


\section{Introduction} \label{sec:intro}

GRB 220101A is an extremely energetic long-duration  gamma-ray burst (GRB) \citep[$T_{90} = 173\pm13$~s;][]{markwardt22gcn} detected on New Year's Day 2022 by \textit{Swift} \citep{tohuvavohu22gcn}. 
Spectroscopic studies determined its redshift to be $z=4.618$ \citep{fynbo22gcn} and in combination with the measured Konus-Wind fluence, this implies an isotropic gamma-ray energy of $E_{\rm iso}\simeq3.6\times10^{54}$ erg \citep{tsvetkova22gcn}, making this burst one of the most energetic ever detected \citep[the 10th-most energetic listed in][]{Burns23}.  
The burst was also detected by the \textit{Fermi}-LAT instrument, and this allowed us to activate a Chandra X-ray Observatory program intended to study the most energetic GRBs with the expectation that these bursts could potentially place limits on the nature of the engine behind the burst and the physics of the afterglow. 
The \textit{Chandra} program also gave us access to the Hubble Space Telescope (\textit{HST}) and the Karl G. Jansky Very Large Array (VLA).  These observatories allow us to extend the study of this burst both in time and in wavelength compared to previous discussions \citep{zhu23} and to follow the afterglow well after the jet break in spite of the afterglow declining as steeply (within the errors) as any other observed long gamma ray burst (LGRB).

In this paper, we analyze the broadband afterglow ($10^4\lesssim\Delta T\lesssim10^7$ s) of the GRB observed in the X-rays (with \textit{Swift}-XRT and \textit{Chandra}), optical-IR (with the \textit{HST} and various ground based telescopes), and submm-radio with the Atacama Large Millimeter Array (ALMA) and the Karl G. Jansky Very Large Array (VLA). In Section 2, we discuss the data analyses and reduction. In Section 3, we conduct a numerical and analytical modelling of the broadband afterglow light curve. In Sections 4 and 5, we discuss our results in the context of GRB physics and how studies like these can open up avenues for better testing of the standard GRB model.

\section{Observations and Data Analysis}

In this section, we present the set of multi-wavelength data obtained, reduced, and analyzed for this work, spanning from soft X-rays to cm-wave radio frequencies.

\subsection{X-rays}
\subsubsection{\textit{Swift} observations}
\textit{Swift} observed the prompt emission of GRB~220101A from 74 to 240s after outburst using the Windowed Timing (WT) mode of its X-ray Telescope (XRT). It returned to the GRB in Photon Counting (PC) mode starting at 3800s after the outburst.  The PC data (along with other data we will discuss) are shown in Figure~1.
The \textit{Swift} data were obtained from the \textit{Swift}/XRT lightcurve repository\footnote{\href{http://www.swift.ac.uk/xrt_curves/}{http://www.swift.ac.uk/xrt\_curves/}} \citep{evans07,evans09}. To convert counts in the full energy range to flux density at 5 keV, we used the parameters of the best \emph{Swift}/XRT spectrum fit: a Galactic neutral hydrogen column density $N_{\mathrm{H,MW}} = 6.29 \times 10^{20}$ cm$^{-2}$, a photon index of $\Gamma_{X,WT} = 1.63$ and a host-galaxy neutral hydrogen column density of $N_{\mathrm{H,int}} = 6 \times 10^{21}$ cm$^{-2}$.\footnote{\url{https://www.swift.ac.uk/xrt_spectra/01091527/}} The conversion to flux density was performed using the web-based Portable Interactive Multi-Mission Simulator ({\sc pimms}\footnote{\href{https://cxc.harvard.edu/toolkit/pimms.jsp}{https://cxc.harvard.edu/toolkit/pimms.jsp}}) tool which gives us a conversion factor of %The conversion factors were $3.512\times10^{-12}$~erg~s$^{-1}$~keV$^{-1}$~cm$^{-2}$~count$^{-1}$ in WT mode and
$3.882\times10^{-12}$~erg~cm$^{-2}$~s$^{-1}$~keV$^{-1}$~count$^{-1}$.

\subsubsection{Chandra Observations}

After about thirteen days, GRB 220101A became too faint for \textit{Swift}/XRT. Therefore, we observed GRB~220101A with the \textit{Chandra} X-ray Observatory in five epochs between February 8 and February 10, 2022 (38 to 40 days after outburst) for a total exposure time of 74,300~s with a mean date of February 9.58 (39.4 days after outburst).   These observations were performed under Proposal 23500404 (PI: Fruchter). 

We obtained a clear detection of the GRB in the combined \textit{Chandra} data. In an aperture of 3 pixels radius, which will contain 90\% to 95\% of the total flux depending on photon energy\footnote{\url{https://cxc.harvard.edu/proposer/POG/html/}}, we detected 17 photons at the location of the GRB.  Based on the measured background we would have expected one photon in the aperture in the absence of a source. Again using the {\sc pimms} calculator with the Swift parameters, and accounting both for the background and aperture correction, we find an unabsorbed flux density of $3.6^{+0.9}_{-0.9} \times 10^{-16} \, {\rm erg} \,{\rm cm}^{-2}\,{\rm s}^{-1}\,{\rm keV}^{-1}$ where the error is dominated by the $\sim 25$\% Poisson error of the source.

\subsection{Optical/IR}

\subsubsection{HST}

\begin{figure*}
    \centering
    \includegraphics[width=\linewidth]{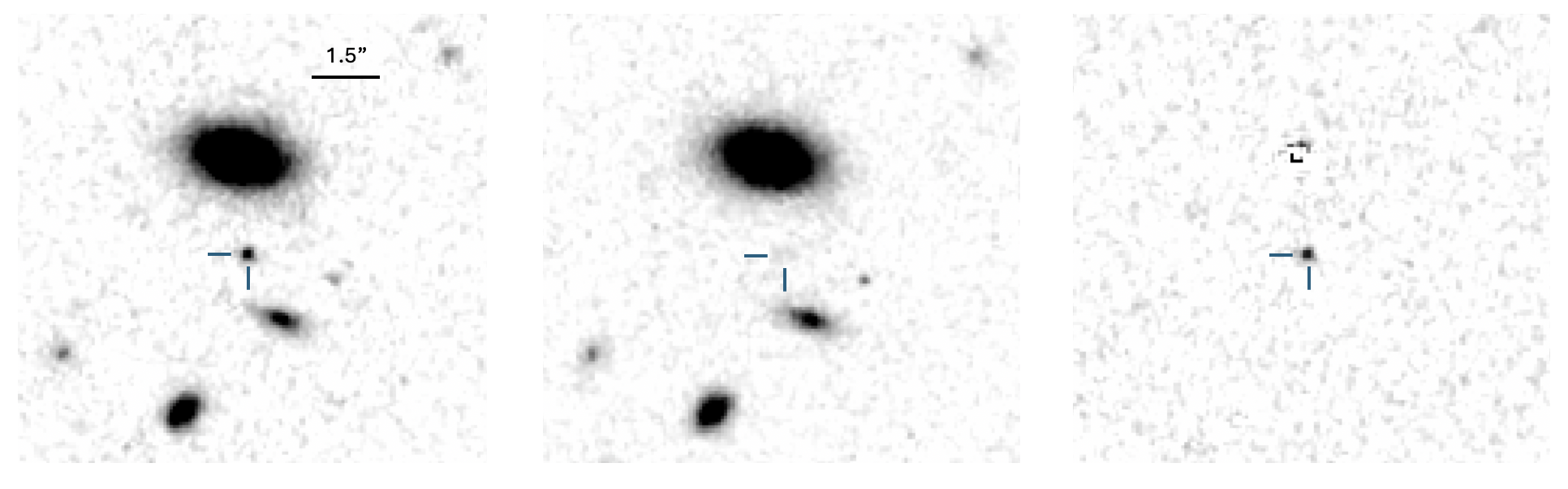}
    \caption{HST F125W image of the GRB 220101A. The two blue lines mark the position of the transient. Left : at $\sim38$ days; Middle: at $\sim236$ days; Right: background subtracted image at $\sim38$ days.}
    \label{fig:hst}
\end{figure*}

We observed the afterglow emission of GRB 220101A with the \textit{HST} under proposal ID 16838 (PI : A. Fruchter), which was awarded through our \textit{Chandra} program. We used the WFC3 in the nIR F125W filter, roughly equivalent to the ground-based $J$ band, and in the nIR/optical F775W, which is close to the ground-based $i$ band. We observed twice, first on February 7, 2022, $\sim$ a month from the burst, and again on August 25, 2022, or nearly 9 months after the burst.  

In both epochs, a single orbit was devoted to each filter, with four exposures taken in four-point dither patterns.   The exposures were combined using \texttt{astrodrizzle} in the \texttt{Drizzlepac} library from STScI onto subsampled output images with pixel scales of $0\farcs02$ in the F775W filter and $0\farcs06$ in the F125W filter.  To obtain the measurement of the optical transient's magnitude at the first epoch without contamination by a possible host, the two epochs were aligned using stars in the field and the second epoch was subtracted from the first.  As the transient was fading as $\sim t^{-3}$ (see Sect. \ref{sec:lightcurve}), any source flux remaining in the second epoch should introduce at most a negligible error in the subtraction. For February 7, 2022 we obtain AB magnitudes of $26.71 \pm 0.13$ in F775W and $25.65 \pm 0.05$ in F125W. The host-galaxy could not be significantly detected due to contamination from two nearby galaxies straddling the GRB (Figure \ref{fig:hst}).

\subsubsection{Ground-based optical/IR data}

Ground-based optical and nIR detections of the afterglow of GRB~220101A were reported by various observers through the NASA General Coordinates Network (GCN)\footnote{\url{https://gcn.nasa.gov/}}. In addition, follow-up data from several different telescopes and not included in the GCNs were published by \citet{zhu23}. However, because of the high redshift of GRB~220101A ($z=4.618$), the rest-frame Ly~$\alpha$ line falls on 6830~\AA. Therefore, the observed optical data in $r/R$ and all bluer bands are affected by the Ly~$\alpha$ forest. This effect on the optical spectrum was demonstrated by \citet{zhu23}. Furthermore, the data at these bands do not include any points after the jet break (see Sect. \ref{sec:lightcurve}). For these reasons, we ignore them in the following analysis and only use data in $i$/F775W and redder bands. We use observations from GCNs 31353 \citep{fu22gcn}, 31357 \citep{perley22gcn}, 31358 \citep{postigo22gcn}, 31361 \citep{vinko22gcn}, 31373 \citep{davanzo22gcn}, 31395 \citep{davanzo22gcn2}, 31401 \citep{guelbenzu22gcn}, 31425 \citep{perley22gcn2}; and all other photometry in these bands included in \citet{zhu23}.

\subsection{Radio : VLA}

GRB220101A was observed by the VLA in 2022 under project ID SN0404 (PI: A. Fruchter) between 13th January and 3rd May and under project ID 21A-405 (PI: T. Laskar) between 4th January and 12th March, through continued monitoring at 5 GHz to 34 GHz. We used the calibrated data directly if available, while for others the raw data was downloaded from the NRAO data archive\footnote{\href{http://data.nrao.edu}http://data.nrao.edu} and processed by the CASA 6.5.4 pipeline. The CASA task \texttt{tclean} (with natural weighting) was run on all the calibrated measurement sets at every observed frequency and the peak flux density (or equivalently the total since it was a point source) of the map was noted. Self-calibration was avoided due to low flux densities ($\lesssim 1$ mJy). Table \ref{tab:radio_fluxes} lists the band/frequency and date for each observation and the observed flux densities and the observed RMS sensitivity in each band.

\begin{table*}
    \centering
    \begin{tabular}{ccccccccc}
    \hline
    \hline
         Obs & Array & Freq. & Project & Date & Phase (t) & $F_{\rm tot}$ & RMS \\
             &      & (GHz) &          & YYYY-MM-DD & (days) & (mJy) & (mJy) \\
        \hline
        VLA & A & 6.0 & 21A-241 & 2022-01-04 & 2.9 &  0.182 & 0.010 \\
        " & " & 6.0 & 21A-241 & 2022-01-08 & 6.9 &  0.322 & 0.009 \\
        " & " & 6.0 & SN0404 & 2022-01-13 & 11.9 &  0.312 & 0.009 \\
        " & " & 6.0 & 21A-241 & 2022-01-14 & 12.9 &  0.201 & 0.011 \\
        " & " & 6.0 & 21A-241 & 2022-01-15 & 13.9 &  0.261 & 0.013 \\
        " & " & 6.0 & SN0404 & 2022-01-22 & 20.9 &  0.220 & 0.010 \\
        " & " & 6.0 & 21A-241 & 2022-01-29 & 27.9 & 0.120 & 0.009 \\
        " & " & 6.0 & SN0404 & 2022-02-08 & 37.9 &  0.060 & 0.007 \\
        " & " & 6.0 & SN0404 & 2022-03-05 & 62.9 &  0.050 & 0.004 \\
        " & " & 6.0 & 21A-241 & 2022-03-12 & 69.9 &  0.039 & 0.009 \\
        \hline
        " & " & 10.0 & " & " & 2.9 &   0.323	& 0.008 \\
        " & " & 10.0 & " & " & 6.9 &   0.458	& 0.009 \\
        " & " & 10.0 & " & " & 11.9 &  0.467	& 0.008 \\
        " & " & 10.0 & " & " & 12.9 &  0.262	& 0.011 \\
        " & " & 10.0 & " & " & 13.9 &  0.300   & 0.011 \\
        " & " & 10.0 & " & " & 20.9 &  0.230	& 0.010 \\
        " & " & 10.0 & " & " & 27.9 &  0.122	& 0.010 \\
        " & " & 10.0 & " & " & 37.9 &  0.056	& 0.005 \\
        " & " & 10.0 & " & " & 62.9 &  0.038	& 0.003 \\
        " & " & 10.0 & " & " & 67.9 &  0.042	& 0.009 \\
        " & " & 10.0 & SN0404 & 2022-04-29 - 2022-03-05 & 123.0 & 0.013 & 0.001 \\
        \hline
        " & " & 15.0 & " & " & 2.9 &   0.447	& 0.009 \\
        " & " & 15.0 & " & " & 6.9 &   0.450	& 0.007 \\
        " & " & 15.0 & " & " & 11.9 & 0.402	& 0.008 \\
        " & " & 15.0 & " & " & 12.9 & 0.277	& 0.011 \\
        " & " & 15.0 & " & " & 13.9 &  0.298	& 0.011 \\
        " & " & 15.0 & " & " & 20.9 &  0.180	& 0.010 \\
        " & " & 15.0 & " & " & 27.9 &  0.134	& 0.011 \\
        \hline
        " & " & 22.0 & " & " & 2.9 &   0.573	& 0.010 \\
        " & " & 22.0 & " & " & 11.9 &  0.317	& 0.009 \\
        " & " & 22.0 & " & " & 12.9 &  0.282	& 0.014 \\
        " & " & 22.0 & " & " & 13.9 &  0.227	& 0.014 \\
        " & " & 22.0 & " & " & 20.9 &  0.180	& 0.011 \\
        " & " & 22.0 & " & " & 27.9 &  0.109	& 0.010 \\
        " & " & 22.0 & " & " & 37.9 &  0.050 & 0.006 \\
        \hline
        " & " & 34.0 & " & " & 2.9 &   0.601	& 0.014 \\
        " & " & 34.0 & " & " & 11.9 &  0.295	& 0.014 \\
        " & " & 34.0 & " & " & 12.9 &  0.245	& 0.020 \\
        " & " & 34.0 & " & " & 13.9 &  0.170	& 0.017 \\
        " & " & 34.0 & " & " & 20.9 &  0.065	& 0.017 \\
        \hline
        ALMA & 12m & 97 GHz & 2021.1.00658.T & 2022-01-02 & 0.9 &   0.810 & 0.012 \\
        ALMA & 12m & 97 GHz & " & 2022-01-04 & 2.9 &   0.552 & 0.014 \\
        ALMA & 12m & 97 GHz & " & 2022-01-08 & 6.9 &   0.310 & 0.017 \\
        ALMA & 12m & 97 GHz & " & 2022-01-22 & 20.9 & 0.122 & 0.018 \\
        \hline
    \end{tabular}
    \caption{Table lists the summary of all radio-submm observations used in this paper: the telescope and array used, the project ID, the observing frequency and date, the total flux density and RMS. The phase column refers to the time elapsed since the burst.}
    \label{tab:radio_fluxes}
\end{table*}

\subsection{Sub-mm : ALMA}

The Atacama Large sub-Millimeter Array (ALMA) observed GRB220101A from 2022-01-02 to 2022-01-22 at Band 3 (97 GHz) under project ID 2021.1.00658.T (PI: T. Laskar). The calibrated data for each project were downloaded from the ALMA data archive\footnote{https://almascience.nrao.edu/} and the task \texttt{tclean} with natural weighting was run on each measurement set. The resulting total flux densities at each frequency were noted for each observation and have been tabulated in Table \ref{tab:radio_fluxes} along with the VLA observations.

\section{Modelling and Analysis}

%In Figure \ref{fig:BPL_fit} we only show the (host) extinction-corrected F814W HST light curves along with the X-ray points from XRT and Chandra. In this section we first aim to understand the physics of this afterglow through simple broken power law modelling of the light curve to get an estimate of the pre and post-break slopes as well as the break timescale. The latter part of this section is devoted to a semi-numerical modelling of the light curve using \texttt{afterglowpy} to investigate the physical parameter space relating to this GRB.

This section aims to investigate the physical nature of the GRB 220101A afterglow through multi-wavelength light curve analysis and modelling using \texttt{afterglowpy} \citep{ryan20} and an analytical fitting code \citep{kangas21} based on \citet{granotsari02}. As per the standard model \citep{mrees99}, GRB afterglow emission is a result of a highly relativistic jet ($\Gamma\gtrsim100$) shocking the CBM to drive a forward shock into the CBM and a reverse shock back into the ejecta to accelerate electrons in both media. The focus of this work is on examining the behavior of the forward shock and, particularly, the geometric jet break. This is a highly time-dependent radiation calculation that includes the time evolution of relevant synchrotron frequencies and the geometric effects of the jet as it decelerates. An afterglow light curve at any given wavelength is generally observed to contain a series of power-law segments between breaks. Each break is observed as a change in light curve slope as the break frequency of the synchrotron emission spectrum moves through the observed frequency. The expected breaks in slope are related to the passing of the self-absorption break (or $\nu_{sa}$), the cooling break ($\nu_c$) and the base synchrotron frequency ($\nu_m$). The sequence of these breaks depends on the external environment, the observing frequency and the physical properties of the GRB. 

In addition to all these breaks, a GRB afterglow can, at late times, display a decay slope that is steeper than the expected slopes from all the synchrotron breaks and which is achromatic across much of the spectrum. This occurs when the continuous deceleration of the jet leads to the relativistic beaming of the light becoming broader than the opening angle of the jet ($\Gamma<1/\theta_j$).  Until this point, while the surface brightness of the jet has fallen as the relativistic beaming angle increased, the area of the jet the observer sees has grown as well.  However, this offset stops abruptly when the relativistic beaming becomes broader than the jet itself, producing a dramatic steepening of the light curve
\citep{mrees99,sari99}.  In addition, lateral expansion of the jet at the speed of sound can further steepen the decline to a post-break slope of $-p$, where $p$ is the power-law slope of the energies of the synchrotron electrons \citep{rhoads99}.  As we will discuss later,  these analytical expectations can be altered by energy injection and hydrodynamic effects. 

\subsection{Light Curve Analysis}
\label{sec:lightcurve}

\begin{figure}
    \centering
    \includegraphics[width=\linewidth]{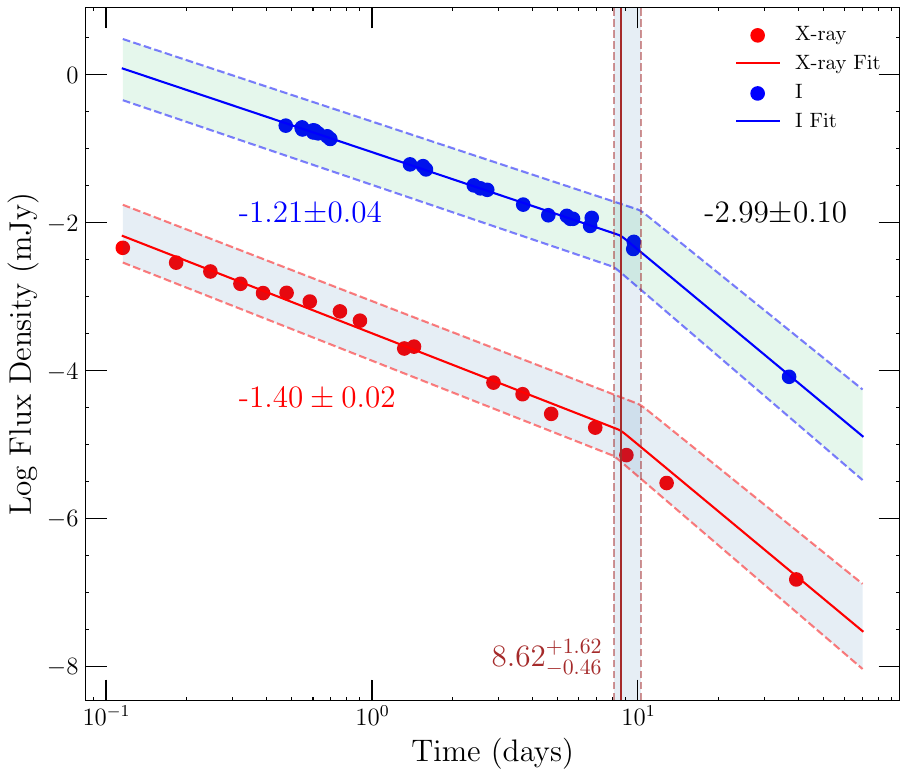}
    \caption{Figure showing the I band and the X-ray points, which were sampled the most among all wavelengths, and the results of the broken power law fit using Equation \ref{eqn:bpl}.}
    \label{fig:BPL_fit}
\end{figure}

Most of the underlying physics in GRB afterglow light curves can be understood by studying the nature of their broadband power law indices and the presence/absence of break(s). For simplicity of discussion, we start by showing in Figure \ref{fig:BPL_fit} only the extinction-corrected ground-based $i$-band and HST F814W observations along with the X-ray data from XRT and Chandra. The X-ray and the optical light curves in Figure \ref{fig:BPL_fit} show a distinct geometric jet break at $\lesssim$ 10 days. We fit the X-ray and optical using a broken power law of the following form:

\begin{equation}
    F_\nu(t)=A\Bigg[\bigg(\frac{t}{t_j}\bigg)^{-\omega\alpha_L}+\bigg(\frac{t}{t_j}\bigg)^{-\omega\alpha_H}\Bigg]^{-1/\omega}
\label{eqn:bpl}
\end{equation}

where $A$ is a normalization parameter, $t_j$ is the break timescale, $\alpha_L$ and $\alpha_H$ are the pre (``low") and post-break (``high") slopes respectively, and $\omega$ denotes the break sharpness, with $\omega=1$ and $\omega\gg1$ implying minimum and maximum sharpness respectively. As the post-jet-break slopes and timescales are expected to be achromatic (for all cases other than $\nu<\nu_m<\nu_c$), we assume the same break parameters for both light curves except the pre-break slope. Figure \ref{fig:BPL_fit} shows the fit with a $\omega=10$, which provides a sharp break. The fit was done using the \texttt{emcee} \citep{foremanmackey13} Markov-chain Monte Carlo (MCMC) package and the resulting best-fit parameters (medians and one sigma values) are displayed in the figure. 

The pre-break slopes for the optical and X-rays are $-1.21\pm0.04$ and $-1.40\pm0.02$, respectively. They differ by $\sim0.19\pm0.05$, whose $1\sigma$ limit, $\sim0.24$, is close to $\Delta\alpha_{\rm X-O}=0.25$, expected when the cooling frequency lies between the optical and the X-rays in a constant-density circumburst medium (CBM) with  $\nu_m<\nu<\nu_c$ or $\nu_m, \nu_c<\nu$ ({\it c.f.} \citealt{granotsari02}). This implies the pre-break indices for the two frequency regimes would be given by $\alpha_X=-(3p_X-2)/4$ and $\alpha_O=-3(p_O-1)/4$ and hence $\Delta\alpha_{\rm X-O}=0.25$, according to theoretical expectations. Plugging the observed values of $\alpha_O$ and $\alpha_X$, we obtain $p=p_O\sim2.61\pm0.03$ and $p=p_X\sim2.86\pm0.05$, whose $1\sigma$ limits are close ($2.64$ and $2.80$ respectively). The post-break slope, in contrast, is $-2.99\pm0.10$. While the break timescale is $\sim8.62$ days as evident from the figure, the post-break slope is markedly steep. The analytic expectation from a jet that expands sideways after the jet break is a post-break index of $p$ implying if the spectral model fits provide a value of $p\sim3.0$, we would be able to confirm lateral expansion. The afterglow models have been discussed in upcoming sections and possibilities of lateral expansion dealt with. The differences between these three post-break slopes ($p_O,p_X$ assuming lateral expansion and $2.99\pm0.10$) are non-negligible, but compatible with the findings of \citet{gompertz18}, who found a scatter of $\sigma_p \approx 0.25$ when obtaining $p$ through different relations for the same GRBs. One should also note that due to the poor sampling of data points around and beyond the jet break, the break timescale and the post-break slopes are correlated and underconstrained. For example, an earlier break time would result in a slightly flatter post-break index and vice versa. It is also possible that we are seeing effects of hydrodynamics that are not captured in usual analytical calculations.

\begin{figure}
    \centering
    \includegraphics[width=\linewidth]{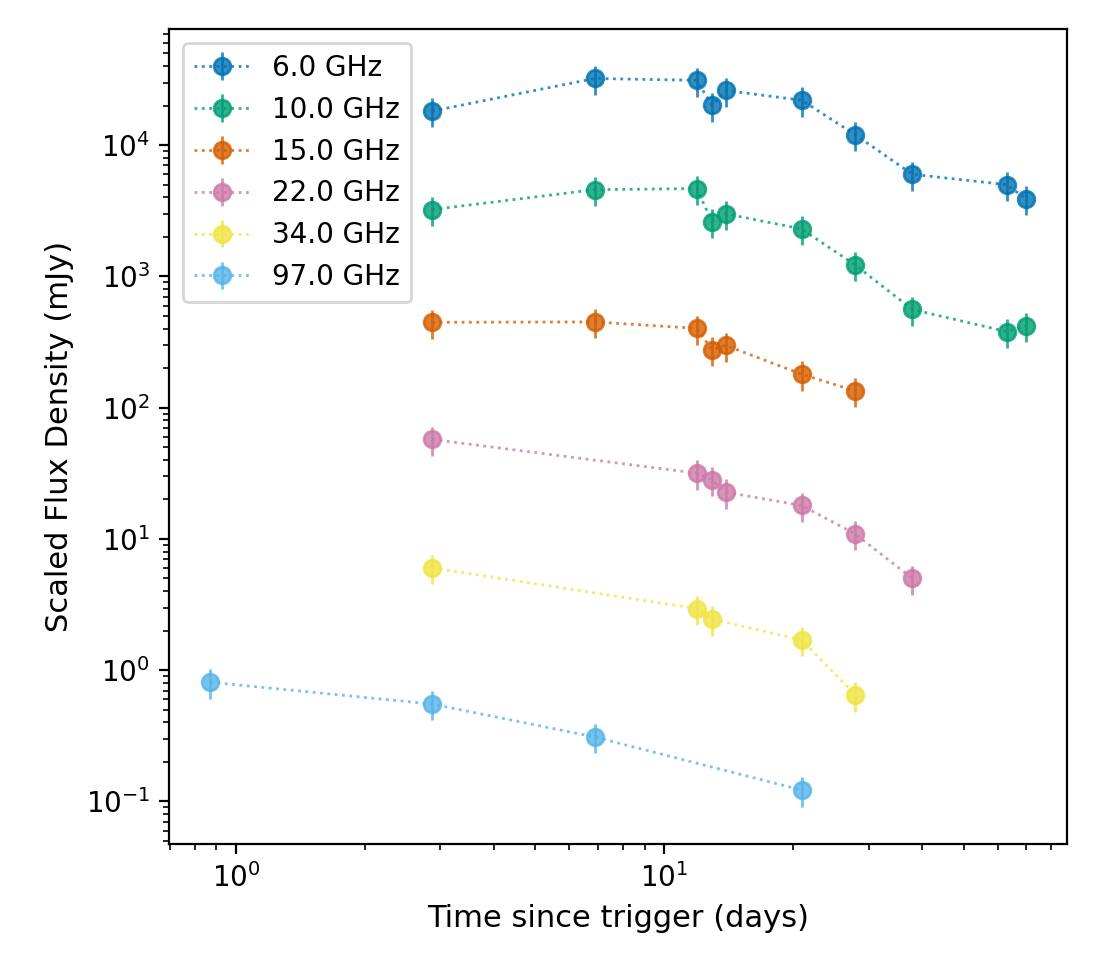}
    \caption{Radio to submm light curves plotted separately (scaled accordingly) for better clarity. While no clear smooth/broken power law behaviour is evident like optical and X-rays, a general change in trend after 10 days is evident for the $6-10$ GHz light curves.} 
    \label{fig:radio}
\end{figure}

The radio frequency points have been plotted separately in Figure \ref{fig:radio} to reduce clutter. While it seems that the radio behaviour does not resemble a broken power law, diffraction and refractive scintillation may cause large enough fluctuations in flux density ($\sim 50\%$) to cause a departure from a smooth power law. There is lack of a clear unifying feature among all frequencies except a possible general steepening after $\sim$ 10 days. It is possible that the peak synchrotron frequency ($\nu_m$) moves through the radio frequencies at $\lesssim10$ days. 

\subsection{Afterglow spectral and light curve modeling}

Afterglows can be modelled either through analytical theory or using sophisticated hydrodynamic simulations. As introduced earlier, \texttt{afterglowpy} is a numerical code that balances both these techniques \citep{ryan20} and which approximates hydrodynamics through calibration to the more rigorous \texttt{BOXFIT} \citep{van12} that includes complete hydrodynamic simulations. To model our light curves, we chose \texttt{afterglowpy} for its ease of use and speed.

While modelling the afterglow, the initial choice of the jet model is important because of viewing angle effects. A structureless, top-hat jet that has a uniform energy profile should produce distinctly different emission patterns from a structured (say, Gaussian) jet that has a different energy cross-section. This effect would be more pronounced if the viewing angle is larger than the jet opening angle, more likely for narrower and faster jets. For this work, we used both a top-hat and a Gaussian jet model in an ISM-like circumburst medium (CBM). Note that both models are only approximations and the jet shape is expected to evolve according to the physics of the GRB engine and the hydrodynamics of the flow through the CBM. 

Since \texttt{afterglowpy} has no internal fitting algorithm, we wrote an MCMC wrapper using \texttt{emcee}, which we gave uniform/log-uniform priors on all relevant parameters (described in Table \ref{tab:priors1}) and tasked it to optimize the $\chi^2$. Since the resulting posterior distributions(s) peaked around the median, we took the median of the resulting posterior distribution of each parameter as the final best-fit value and the 16th and 84th percentiles as one-sigma errors. The free parameters included the total isotropic-equivalent kinetic energy $E_{\rm iso}$, the electron energy distribution (EED) spectral index $p$, the viewing and opening angles $\theta_{\rm obs}$ and $\theta_j$, the CBM density $n_0$ and the fraction of internal energy in electrons and magnetic fields as $\epsilon_e$ and $\epsilon_B$ respectively. We added host-galaxy extinction correction through $A_B$ using the extinction laws of \cite{pei92}. We ran \texttt{afterglowpy} for two cases of the electron participation fraction $\xi$ (the fraction of thermal electrons that are accelerated to non-thermal energies and thus take part in the synchrotron radiation) at $0.1$ and $1.0$, where respectively 10\% and 100\% of the electrons radiate. While for a large enough number of MCMC steps (8000 here) and walkers (800 here) the choice of the initial conditions is immaterial, we provided an initial prior based on a \textit{trial and error} \texttt{afterglowpy} fit to the light curve and thereafter used the median of the initial posterior as the initial values for all subsequent MCMC runs for the source.

For each fit, we binned the X-ray light curve, taking the average epoch and flux in each time bin of 0.1~dex starting from the first PC-mode observation. We furthermore added a 10 per cent error floor in the X-ray and optical light curves and 25 per cent in the radio. This was done in order to reduce the overweighting produced by the large number of \textit{Swift} X-ray points, and to prioritize a good fit in the late-time optical and X-ray over the radio (see Sect. \ref{sec:disc}). Additionally, the radio de-weighting helps account for the scatter in radio points caused by refractive and diffractive scintillation. %\km{table captions can be kept either at the beginning or end of the table} %, and the X-ray binning also has the benefit of reducing the time required for fitting.

\begin{table}
\centering
\begin{small}
\begin{tabular}{llc}
\hline
Parameter (unit) & Prior & Range\\
\hline
$E_{\rm K,iso}$ (erg) & Log-uniform & [52.0:57.0] \\
$p$ & Uniform & [2.0:3.0] \\
$n_0$ (cm$^{-3}$) & Uniform & [$10^{-6}$:2.0] \\
$\theta_j$ (rad) & Uniform & [$10^{-4}$:0.5] \\
$\theta_{\rm obs}$ (rad) & Uniform & [$10^{-4}$:0.5] \\
$\epsilon_e$ & Uniform & [$10^{-5}$:$1$] \\
$\epsilon_B$ & Uniform & [$10^{-5}$:$1$] \\
$A_B$ (mag) & Uniform & [0:5] \\
 \hline
\end{tabular}
\caption{Parameters and priors of our \texttt{afterglowpy} fits.}
\label{tab:priors1}
\end{small}
\end{table}

\subsubsection{Top-hat Jet}

A top-hat jet is a simple jet model frequently used in simulations/numerical calculations. It has a uniform energy across its cross section and zero beyond, that is $E(\theta)=E(\theta=0)$ for $\theta
\leq\theta_j$ and $E(\theta)=0$ for $\theta>\theta_j$, where $\theta$ is the angle between the jet axis and a line drawn to a piece of the jet from the bottom of the jet axis, and $\theta_j$ denotes the jet-edge.

Table \ref{tab:pars_tophat} lists the median best-fit parameters for the Top-Hat jet from the MCMC fit for each of the two participation fractions. While it is expected that with higher electron participation fractions both the total required energy and the required CBM density (or the CBM mass that needs to be shocked keeping the same volume) will drop, we find that for $\xi=1.0$ the CBM density drops extremely low, to $\sim10^{-5}$ cm$^{-3}$. ISM density posteriors lower than $10^{-4}$ cm$^{-3}$ are, perhaps surprisingly, not uncommon when fitting multi-wavelength afterglows \citep[e.g.,][]{kangas20,kangas21,cunningham20} but it is unclear how physical such a value is. \cite{zhu23} obtained different best-fit parameters though, and in Appendix \ref{app:zhu} we try to bridge the differences in the best-fits between \cite{zhu23} and ourselves. %For rare situations, low ISM densities $\sim10^{-3}$ cm$^{-3}$ may be associated with galactic halos that may harbor short GRBs \citep[e.g.,][]{debarros11}. With lower $\xi$, as in as low as $0.01$, the CBM density reaches $0.01$ cm$^{-3}$, which implies $n_0$ should range between $10^{-5}$ and $0.01$ cm$^{-3}$.
\iffalse
\begin{table}
    \centering
    \footnotesize
    \begin{tabularx}{0.375\textwidth}{c|cc}
    \hline
    Parameter &  $\xi=0.1$ & $\xi=1.0$ \\
 \hline
  $\log\,E_{\rm K, iso}$ (erg)   & $56.40^{+0.07}_{-0.10}$ & $56.17^{+0.07}_{-0.05}$  \\
  p & $2.01\pm0.01$ & $2.16\pm 0.02$ \\
  $\log\,\theta_{\rm obs}$ (rad) & $-2.29\pm 0.05$ & $-2.45\pm 0.04$ \\
  $\log\,\theta_j$ (rad)  & $-2.03\pm0.05$ & $-2.19\pm0.03$ \\
  $\log\,n_0$ (cm$^{-3}$) &$-3.47\pm^{+0.40}_{-0.37}$ & $-5.44^{+0.25}_{-0.33}$\\
  $\log\,\epsilon_e$ & $-0.37^{+0.16}_{-0.18}$ & $-1.38^{+0.11}_{-0.09}$ \\
  $\log\,\epsilon_B$ & $-1.09^{+0.42}_{-0.47}$ & $-1.84^{+0.34}_{-0.31}$ \\
  $A_B$ & $-0.59\pm0.01$ & $0.13^{+0.04}_{-0.02}$ \\
  \hline
  $E_{\rm K}$ (erg) & $3.3\times10^{51}$ & $3.1\times10^{51}$ \\
  $\chi^2$/DOF & 8.80 & 2.12 \\
  \hline
    \end{tabularx}
   \caption{Best-fit parameters for the Tophat jet for different values of $\xi$.}
    \label{tab:pars_tophat}
\end{table}
\fi

\begin{table}
    \centering
    \footnotesize
    \begin{tabularx}{0.375\textwidth}{c|cc}
    \hline
    Parameter & $\xi=0.1$ & $\xi=1.0$ \\
    \hline
    $\log (E_{\rm K, iso} / \text{erg})$ & $56.28^{+0.28}_{-0.31}$ & $56.17^{+0.07}_{-0.05}$ \\
    $p$ & $2.02\pm0.01$ & $2.16\pm 0.02$ \\
    $\log (\theta_{\rm obs} / \text{rad})$ & $-2.22^{+0.04}_{-0.05}$ & $-2.45\pm 0.04$ \\
    $\log (\theta_j / \text{rad})$ & $-1.98^{+0.06}_{-0.04}$ & $-2.19\pm0.03$ \\
    $\log (n_0 / \text{cm}^{-3})$ & $-2.39^{+0.50}_{-0.70}$ & $-5.44^{+0.25}_{-0.33}$\\
    $\log \epsilon_e$ & $-0.20^{+0.18}_{-0.35}$ & $-1.38^{+0.11}_{-0.09}$ \\
    $\log \epsilon_B$ & $-1.72^{+0.56}_{-0.29}$ & $-1.84^{+0.34}_{-0.31}$ \\
    $A_B$ (mag) & $0.46^{+0.15}_{-0.21}$ & $0.13^{+0.04}_{-0.02}$ \\
    \hline
    $E_{\rm K}$ (erg) & $1.0\times10^{52}$ & $3.1\times10^{51}$ \\
    $\chi^2$/DOF & 8.80 & 2.12 \\
    \hline
    \end{tabularx}
    \caption{Best-fit parameters for the Top-hat jet for different values of $\xi$.}
    \label{tab:pars_tophat}
\end{table}

\begin{figure}
\label{fig:xi_compare}
\vbox
{
    \includegraphics[width=\linewidth]
    {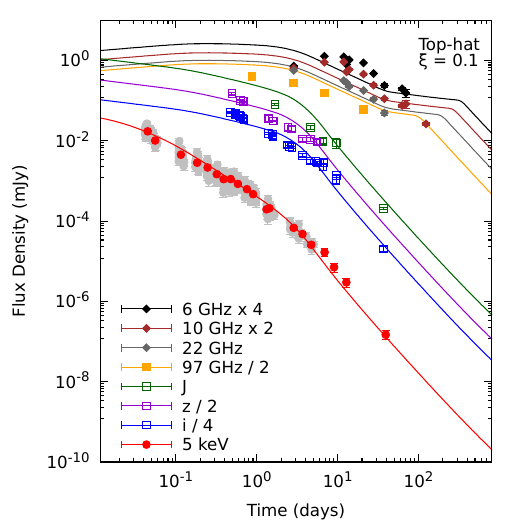} 
    \includegraphics[width=\linewidth]
    {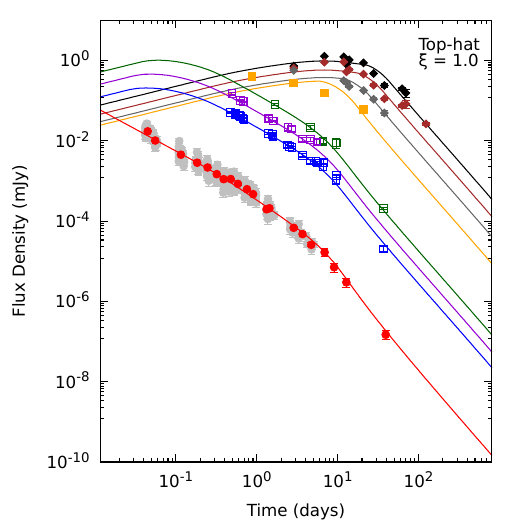}
}
\caption{\texttt{Afterglowpy} top-hat jet fits to the light curves for $\xi=0.1$ and $\xi=1.0$. Both the optical/IR and the radio data are fit better with the higher $\xi$. The grey points show the unbinned X-ray data.}
\label{fig:tophat_lc}
\end{figure}

Figure \ref{fig:tophat_lc} shows the resulting light curve fits for $\xi=0.1$ and $1.0$. While the X-ray observations are well fit by our model irrespective of $\xi$, the behaviour of both the optical/IR and the radio fit depends on $\xi$. Only at high $\xi$ do we see a general similarity in tendency between the data and the model, more so for the radio. %This is exemplified in Figure \ref{fig:lc_only_radio}, where for $\xi=0.1$ the model misses the observed radio trend, but follows it for $\xi=1.0$, even though the fit is not ideal. 
For the optical/IR, even the pre-break light curve decay is only captured by the model at $\xi=1.0$. %Increasing $\xi$ to 1.0 lowers the predicted CBM density, which in turn allows a wider jet width (or the opening angle) as the energy requirements become minimal as the entire isotropic equivalent energy can now be used for particle acceleration and non-thermal radiation.
A higher $\xi$ also produces a larger jet opening angle, and the jet-break timescale is very sensitive to the opening angle: $t_B\propto(E_{\rm K, iso}/n_0)^{1/3}\theta_j^{8/3}$ \citep{sari99}. As evident in Table \ref{tab:pars_tophat} and \ref{fig:xi_compare}, the larger opening angle has shifted the predicted break timescale to lower values, thereby fitting the radio better. Accordingly, the lowest chi-squared we obtain is for $\xi=1.0$, as shown in Table \ref{tab:pars_tophat}.

\begin{figure}
\vbox
{
    \includegraphics[width=\linewidth]
    {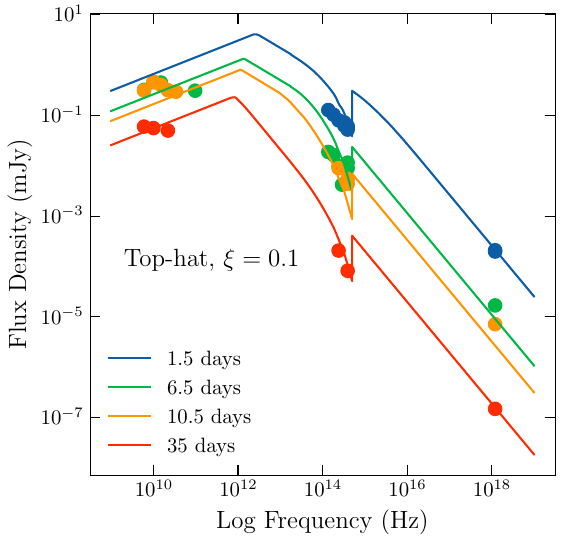} 
    \includegraphics[width=\linewidth]
    {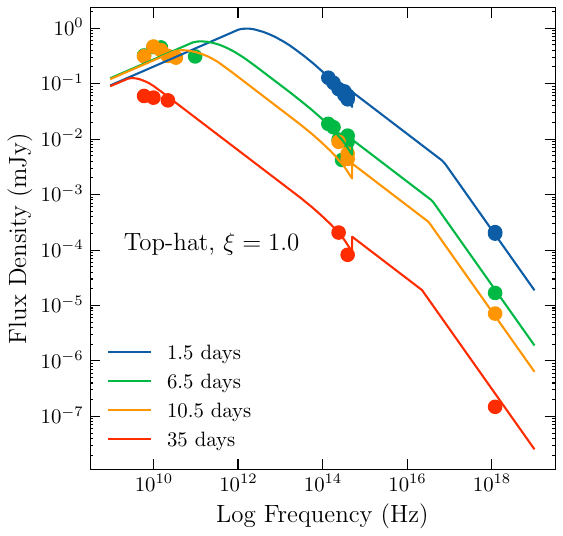} 
}
\caption{Afterglowpy spectral evolution for $\xi=0.1$ and $\xi=1.0$ with the top-hat jet. The evolution of the break frequencies is considerably different between the two models, and the $\xi=1.0$ model fits the X-ray evolution better. The kink in the figures is due to the Heaviside-like (step function in frequency) extinction correction applied to the optical bands.}
\label{fig:tophat_spec}
\end{figure}

\iffalse
Given the best-fit results, it is imperative to consider the relevant timescales involved. Consider below the timescale where the cooling frequency passes the observing frequency \citep{sari98},

\begin{equation}
t_c=0.2\epsilon_B^{-3}E^{-1}_{52}n_0^{-1}\nu^{-2}_{15}\,{\rm \,hours}~.
\end{equation}

which equals $\sim$ and $\sim$ for $\xi=0.1$ and $\xi=1.0$.

\fi

The cooling frequency $\nu_c$ is different at different values of $\xi$, as evident from our Top-hat model spectra shown in Figure \ref{fig:tophat_spec}. The kink in the figure is due to the Heaviside-like extinction correction applied to the optical bands. $\nu_c$ lies between the optical and the X-rays for the entire duration of the afterglow at $\xi=1.0$. This is consistent with the power-law slopes shown in Fig. \ref{fig:BPL_fit}. However, for $\xi=0.1$, we observe that the cooling frequency stays less than the optical frequencies ($\nu_c \lesssim 10^{14}$ Hz) after $\sim5$ days evident in Figure \ref{fig:tophat_spec}.

%\textcolor{blue}{This turns out to be $t_c\simeq$ for the optical (using $\nu=10^{15} Hz)$}

We also note that \texttt{afterglowpy} does not incorporate synchrotron self-absorption as of yet in the latest version, due to the fact that $\nu_\mathrm{sa}$ is generally much lower than the observed radio frequencies during most of the afterglow emission (\textbf{pvt comm.}). For example, the self-absorption frequency for slow cooling ($\nu_c>\nu_m$) in a constant density CBM is given as \citep{granot00} :
\begin{equation}
\nu_{\rm sa}=1.24\times10^9(1+z)^{-1}\epsilon_e^{-1}\epsilon_B^{1/5}n_0^{3/5}E^{1/5}_{52}\frac{(p-1)^{3/5}}{(3p+2)^{3/5}}\,{\rm Hz}
\end{equation}

Plugging in values from Table \ref{tab:pars_tophat} at $\xi=0.1$, $\nu_{\rm sa}\simeq6$ MHz; estimated values are even lower for higher $\xi$. However, in the case of a burst with a high surrounding ISM density, at early times one might expect some lower-energy radio frequencies to be below $\nu_{\rm sa}$, where self-absorption would be important.

%Among the observed frequencies, it is only in the GHz frequencies that one can observe the complete passing of $\nu_m$ and $\nu_c$ starting from $\nu<\nu_m<\nu_c$ to $\nu_m,\nu_c<\nu$ through the entire afterglow time period. Hence the radio light curve behaviour generally looks different from the optical and X-rays, with possibly more spectral breaks than the former.

\iffalse

Figure \ref{fig:lc_only_radio} shows the top-hat jet fit behaviour for the radio flux densities. For $\xi=1.0$, the model captures the trend in the observed radio light curves.

\begin{figure}
\vbox
{
    \includegraphics[width=\linewidth]
    {for_paper_tophat_radio_xi0.1.pdf} 
    \includegraphics[width=\linewidth]
    {for_paper_tophat_radio_xi1.0.pdf} 
}
\caption{Figure showing the \texttt{afterglowpy} fit for the entire submm-radio data only for $\xi=0.1$ and $\xi=1.0$. Clearly the higher $\xi$ model captures the radio behaviour much better.}
\label{fig:lc_only_radio}
\end{figure}

\fi

\subsubsection{Gaussian Jet}

A Gaussian jet is a structured jet, implying its energy varies across the jet and is a Gaussian function of $\theta$. This is given as $E(\theta)=E_{\rm iso}e^{-\theta^2/2\theta_j^2}$ for $\theta\leq\theta_W$ and $E(\theta)=0$ otherwise, where $\theta_W$ is a defined jet edge, or where the Gaussian wing is truncated. Here we choose $\theta_W=4\theta_j$ which has been used typically in literature, generally to be the maximum width of the jet (see e.g., van Eerten et al. 2024).

Table \ref{tab:pars_gaussian_updated} lists the median best-fit parameters for the Gaussian jet from the MCMC fit for each of the participation fractions listed. Most of the parameters are similar to those of the top-hat jet model.
%, except the high $\xi$ case of the opening angle, which is $\sim$ an order of magnitude higher than that of the top-hat jet.

\iffalse
\begin{table}
    \centering
    \footnotesize
    \begin{tabularx}{0.375\textwidth}{c|cc}
    \hline
    Parameter & $\xi=0.1$ & $\xi=1.0$ \\
 \hline
  $E_{\rm K, iso}$ (erg) & $2.5^{+0.6}_{-0.6}\times10^{56}$ & $2.0^{+0.7}_{-0.6}\times10^{56}$  \\
  p & $2.02^{+0.02}_{-0.01}$ & $2.08^{+0.03}_{-0.02}$ \\
  $\theta_{\rm obs}$ (rad) & $0.007^{+0.002}_{-0.003}$ & $0.004\pm0.001$ \\
  $\theta_j$ (rad) & $0.008\pm0.001$ & $0.0060^{+0.0010}_{-0.0006}$ \\
  $n_0$ (cm$^{-3}$) & $0.002\pm0.001$ & $2.46^{+2.85}_{-0.92}\times10^{-5}$\\
  $\epsilon_e$ & $0.18^{+0.11}_{-0.06}$ & $0.13^{+0.05}_{-0.04}$\\
  $\epsilon_B$ & $0.02^{+0.03}_{-0.01}$ & $0.004^{+0.002}_{-0.001}$\\

  \hline
    \end{tabularx}
    \caption{Best-fit parameters for the Gaussian jet for different values of $\xi$.}
    \label{tab:pars_gaussian}
\end{table}
\fi

\begin{table}
    \centering
    \begin{tabularx}{0.375\textwidth}{c|cc}
    \hline
    Parameter & $\xi=0.1$ & $\xi=1.0$ \\
    \hline
    $\log  (E_{\rm K, iso} / \text{erg})$ & $56.21_{-0.12}^{+0.19}$ & $56.36_{-1.03}^{+0.11}$ \\
    $p$ & $2.01_{-0.00}^{+0.00}$ & $2.12_{-0.08}^{+0.05}$ \\
    $\log  (\theta_{\rm obs} / \text{rad})$ & $-2.19_{-0.04}^{+0.03}$ & $-2.32_{-0.04}^{+0.24}$ \\
    $\log  (\theta_{c} / \text{rad})$ & $-2.18_{-0.04}^{+0.04}$ & $-2.28_{-0.04}^{+0.16}$ \\
    $\log  (n_{0} / \text{cm}^{-3})$ & $-2.61_{-0.39}^{+0.35}$ & $-4.93_{-0.06}^{+1.26}$ \\
    $\log  \epsilon_e$ & $-0.04_{-0.07}^{+0.07}$ & $-1.12_{-0.12}^{+1.11}$ \\
    $\log  \epsilon_B$ & $-0.76_{-0.74}^{+0.58}$ & $-2.20_{-0.32}^{+2.12}$ \\
    $A_B$ (mag) & $0.37_{-0.02}^{+0.03}$ & $0.20_{-0.06}^{+0.41}$ \\
    \hline
    $E_{\rm K}$ (erg) & $3.5\times10^{51}$ & $3.2\times10^{51}$ \\
    $\chi^2$/DOF & 8.56 & 4.77 \\
    \hline
    \end{tabularx}
    \caption{Best-fit parameters for the Gaussian jet for different values of $\xi$, with values taken from the provided corner plots.}
    \label{tab:pars_gaussian_updated}
\end{table}

\begin{figure}
\vbox
{
    \includegraphics[width=\linewidth]
    {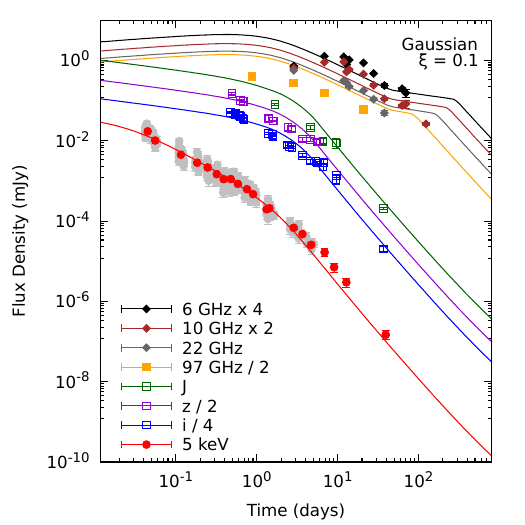} 
    \includegraphics[width=\linewidth]
    {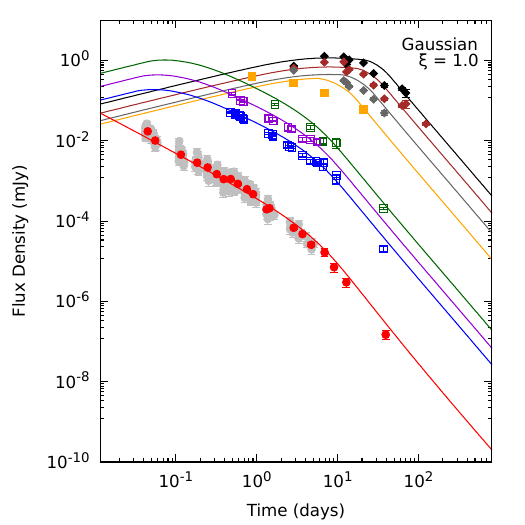}  
}
\caption{\texttt{Afterglowpy} Gaussian jet fits to the light curves for $\xi=0.1$ and $\xi=1.0$. Both the optical/IR and the radio data are fit better with the higher $\xi$.}
\label{fig:gaussian_lc}
\end{figure}

Figure \ref{fig:gaussian_lc} shows the resulting light curve fits for $\xi=0.1$ and 1.0. The fit behaviour is very similar to the Top-hat case, implying the fit results are not sensitive to the presence or absence of a structured jet.

\subsubsection{Fit excluding radio observations and the case of GRB 160625B}

\begin{figure}
    \centering
    \includegraphics[width=\linewidth]{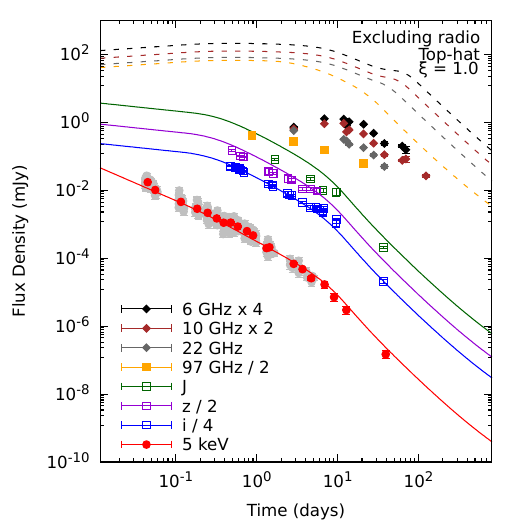} 
    \caption{\texttt{Afterglowpy} top-hat jet $\xi=1.0$ light curve without using radio (dashed lines) in the fit.}
    \label{fig:noradio}
\end{figure}

Here we investigate to what extent the resulting GRB parameters for GRB220101A are driven by the radio behavior. Figure \ref{fig:noradio} shows the resulting plot from only fitting the X-rays and the optical. The expected radio emission is almost an order of magnitude higher than what is observed when $\xi = 1.0$, despite fitting the observations fairly well when they are included in the fit. Further, Table \ref{tab:pars_noradio} shows the resulting best-fit parameters at $\xi=1.0$ for a top-hat jet when radio points are excluded from the fit. Interestingly, the predicted ISM density is much higher, $\sim100$ times larger, when the radio is neglected. 

\begin{table}
    \centering
    \footnotesize
    \begin{tabularx}{0.25\textwidth}{c|c}
    \hline
    Parameter & $\xi=1.0$ \\
    \hline
    $\log (E_{\rm K, iso} / \text{erg})$ & $56.53^{+0.08}_{-0.23}$ \\
    $p$ & $2.04^{+0.01}_{-0.01}$ \\
    $\log (\theta_{\rm obs} / \text{rad})$ & $-1.83^{+0.08}_{-0.08}$ \\
    $\log (\theta_j / \text{rad})$ & $-1.62^{+0.08}_{-0.09}$ \\
    $\log (n_0 / \text{cm}^{-3})$ & $-0.44^{+0.50}_{-0.52}$ \\
    $\log \epsilon_e$ & $-0.02^{+0.01}_{-0.01}$ \\
    $\log \epsilon_B$ & $-3.37^{+0.50}_{-0.44}$ \\
    $A_B$ (mag) & $0.71^{+0.01}_{-0.01}$ \\
    \hline
    \end{tabularx}
    \caption{Best-fit parameters for the Top-hat jet fits ignoring or extremely underweighting radio observations.}
    \label{tab:pars_noradio}
\end{table}

The radio afterglow of GRB~160625B has shown similar behavior. \cite{kangas20} and \citet{kangas21} noted that this GRB's radio emission differed from predictions of the standard model, and numerical and analytical models failed to reproduce the late-time radio behavior. In addition, its CBM density was noted to also be on the order of $10^{-5}$~cm$^{-3}$ \citep{cunningham20,kangas20}. In Appendix \ref{app:160625b}, we include a test similar to the above, and show that removing the radio from the fit of GRB~160125B increases the estimated CBM density by more than an order of magnitude. 
\subsection{Analytical fit}

\begin{figure}[h!]
\centering
\includegraphics[width=\linewidth]{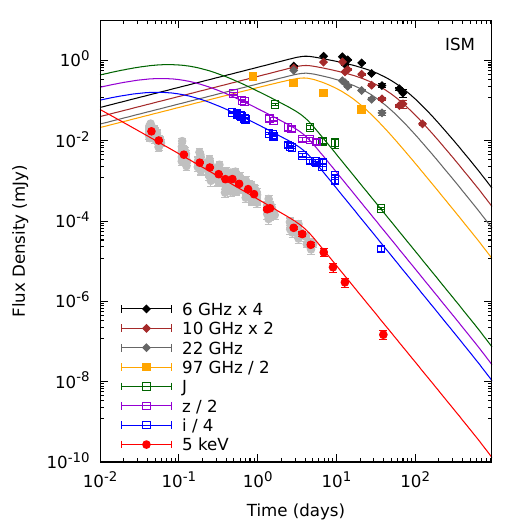} \\
\includegraphics[width=\linewidth]{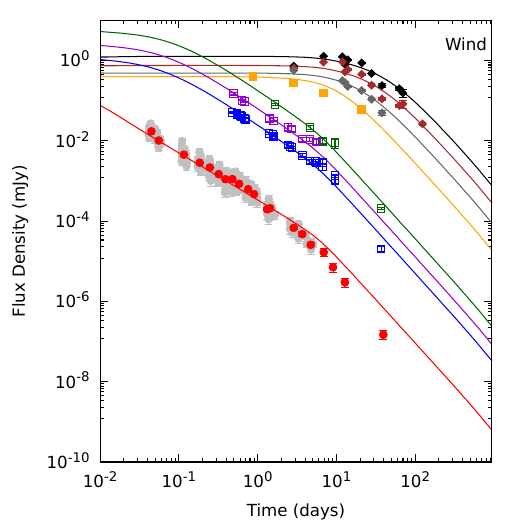}
\caption{Analytical fits to the light curves for ISM- (top) and wind-type (bottom) CBM. The ISM fit is better at replicating the post-break slope.}
\label{fig:gerbillc}
\end{figure}

Since \texttt{afterglowpy} lacks a wind-type CBM treatment and does not incorporate synchrotron self-absorption \citep{ryan20}, we also used an analytical code implementing the spectrum of a top-hat jet afterglow in ISM- and wind-type CBM \citep{granotsari02}. This custom \texttt{python}-based MCMC code was originally written for the analysis of a sample of multi-wavelength afterglows in \citet{kangas21}. The code uses the \texttt{emcee} package and includes implementations of inverse Compton radiation \citep{sariesin01} and jet breaks with and without lateral expansion \citep{mrees99,rhoads99}. For a more detailed description of the modeling code, see \citet{kangas21}. The free parameters, similar to those of the \texttt{afterglowpy} fit, include $\theta_\mathrm{j}$, $E_\mathrm{K,iso}$, $n_0$ (ISM-type CBM) or the dimensionless density parameter $A_*$ (wind-type CBM), $p$, $\epsilon_e$, $\epsilon_B$ and $A_B$. We again adopted wide, uniform or log-uniform prior ranges for each parameter, with 10 or 25 per cent error floors in optical/X-ray and radio bands respectively. The parameters and their prior ranges are listed in Table \ref{tab:priors1}, while the median posterior parameters and their uncertainties are listed in Table \ref{tab:gerbil_results}. The best fits are shown in Fig. \ref{fig:gerbillc}. 

The wind model fails to replicate the late-time decline. The best ISM fit is better in terms of $\chi^2$, and matches the observations in the radio and optical reasonably well, while the X-ray decline by eye is slightly too shallow both before and after the break. This is because the model puts $\nu_c$ above 5~keV until the last observed epoch, while our power-law fits indicate this is not the case (see Sect. \ref{sec:lightcurve}). Additionally, the best ISM fit again includes an extremely low density -- on the same order as the \texttt{afterglowpy} fit with $\xi = 1.0$ described above. Both fits additionally have an extremely narrow jet opening angle at the edge of the prior.

\begin{table}[h!]
\centering
\begin{small}
\begin{tabularx}{0.375\textwidth}{c|cc}
\hline
Parameter (unit) & Prior & Range\\
\hline
$\theta_\mathrm{j}$ (rad) & Log-uniform & [0.01:0.5] \\
$E_\mathrm{K,iso}$ (erg) & Log-uniform & [$10^{49}$:$10^{57}$] \\
$n_0$ (cm$^{-3}$) or $A_*$ & Log-uniform & [$10^{-6}$:$10^{5}$] \\
$p$ & Uniform & [2.0:3.2] \\
$\epsilon_e$ & Log-uniform & [$10^{-5}$:$1$] \\
$\epsilon_B$ & Log-uniform & [$10^{-5}$:$1$] \\
$A_B$ (mag) & Uniform & [0:5] \\
 \hline

\end{tabularx}
\caption{Parameters and priors of our analytical afterglow fits.}
\label{tab:priors1}
\end{small}
\end{table}

\begin{table}[h!]
\centering
\begin{small}
\begin{tabularx}{0.46\textwidth}{c|cc}
\hline
Parameter (unit) & ISM & Wind \\
\hline
log ($E_\mathrm{K,iso}$/erg) & $56.04^{+0.07}_{-0.11}$ & $55.47\pm0.04$ \\
$p$ & $2.41^{+0.02}_{-0.03}$ & $2.18\pm0.02$ \\
log ($\theta_\mathrm{j}$/rad) & $-1.994^{+0.009}_{-0.002}$ & $-1.997^{+0.006}_{-0.003}$ \\
log ($n_0$/cm$^{-3}$) or log $A_*$ & $-5.16^{+0.16}_{-0.07}$ & $-1.67\pm0.04$ \\
log $\epsilon_e$ & $-1.54^{+0.06}_{-0.03}$ & $-1.44\pm0.02$\\
log $\epsilon_B$ & $-2.02^{+0.23}_{-0.20}$ & $-1.33\pm0.10$ \\
$A_B$ (mag) & $0.13\pm0.02$ & $0.19\pm0.02$ \\
 \hline
$\chi^2$/DOF & 4.6 & 5.8 \\
 \hline
\end{tabularx}
\caption{Parameter posterior distributions in our best analytical afterglow fits.}
\label{tab:gerbil_results}
\end{small}
\end{table}

\section{Discussion}
\label{sec:disc}

\subsection{Nature of the post-break light curve : sideways expansion or a geometric effect?}

In the absence of instabilities, a collimated conical jet is confined by the external medium, with strict pressure boundary conditions defined along its boundary \citep[e.g.,][]{daly88}. This defines the typical jet ``size" and its opening angle $\theta_j$. As the jet decelerates by interaction with its medium, the luminosity typically decreases. An achromatic break in the jet light curve arises when the size of the beaming cone exceeds that permitted by the size of the jet, or the beaming angle exceeds the jet opening angle ($1/\Gamma>\theta_j$), after which the rate of decrease in luminosity is even faster since now in addition to the decrease in $\Gamma$, the observed luminosity is increasingly a smaller fraction ($s_\Omega/s_\Gamma$, where $s$ refers to the angular size of the relevant cone and $s_\Gamma$ increases in time to a maximum of $2\pi(1-\cos1/\Gamma)$) of the total theoretical luminosity \citep[e.g.][]{kumarzhang15}. When the jet is ultra-relativistic, or $\Gamma\gg1/\theta_j$, there is not enough time for the jet in its own frame to expand sideways. However, during the jet break, or where $\Gamma\sim1/\theta_j$, the assumption breaks down and the jet is free to expand laterally. If this expansion proceeds at the sound speed, one should expect two breaks in the light curve, one that is geometric where $\Gamma\sim1/\theta_j$, and the other for sideways expansion when $\Gamma$ drops to $1/(\sqrt{3}\theta_j)$ \citep{granot99,granot00,granot01,granotsari02, piran04}. Note that for a structured jet, this will change. In contrast, if the expansion is at the speed of light, both the breaks would coincide. However, one would need extremely fine light curve sampling to observe an effect such as this, which we do not have here, and the breaks will not be infinitely sharp, so they will be combined into one. Lateral expansion would result in an even further reduction in luminosity toward our line of sight due to expansion of the jet material itself \citep[e.g.][]{rhoads99}. %\asfp{This paragraph needs (a) reference(s)}

In the case of lateral expansion at an observing frequency $\nu>\nu_c>\nu_m$ or $\nu_c>\nu>\nu_m$, which is true for the post-break X-ray and optical light curves in this paper respectively (Figure \ref{fig:tophat_spec}), the expected post-break slope is $p$. From simple pre-break broken power law fits, numerical and analytical modelling, we could determine $p$ to be in the range of 2.00-2.60. However, the \textit{observed} post-break achromatic slope is $-2.99\pm0.10$ (Figure \ref{fig:BPL_fit}) -- higher at least by 6-7$\sigma$ than the $p$ from model fits. Such steep slopes are relatively rare, although have been observed in few GRBs \citep[e.g.,][]{wang18,panai07} but with much larger errors ($\sim 0.5$ in slope) making the detection much less certain. This is clearly not an obvious example of the kind of lateral expansion predicted by \cite{rhoads99}, which is not predicted to proceed this fast. Indeed, \cite{zhang09} show through hydrodynamic simulations that in many cases it may be incorrect to expect a simple $p$ decay in the case of lateral expansion due to a number of more complex effects. 

One of the primary conclusions from \cite{zhang09} remains that numerical simulations show that lateral expansion is much slower than what one expects from simple analytic theory as in \cite{rhoads99}, \cite{sari99} or \cite{piran04}. This partly stems from the fact that the jet ceases to be structureless (top-hat) at $\Gamma\sim1/\theta_j$ and as it expands sideways. For an almost on-axis viewer (as we believe is the case for GRB 220101A from fitting models), the inclusion of time-lag effects from the off-axis would make the outer jet brighter than the inner as the off-axis radiation comes from a time when the jet was faster/had a higher energy density. This would create a limb-brightened jet \citep{granot99,zhang09}. A transition from a highly relativistic regime to a regime where $\Gamma<1/\theta_j$ would hence be accompanied by a jet break but the post-break slope would strongly depend on the jet energy, viewing angle and the opening angle. This post-break slope would be independent of the presence of lateral expansion. It is hence clear that the analytical theory of lateral expansion agrees with numerical simulations to only a certain extent as discussed in e.g., \cite{zhang09}, \cite{vaneerten11} or \cite{granot01}.

If the post-break slope was shallower than $p$, lateral expansion could have certainly been ruled out. However, since it is much steeper, lateral expansion cannot be ruled out, while the geometric edge effect alone can. Since a GRB is a rare event that does not ``repeat", we must make sure proper monitoring exists especially at late times for future GRBs, so that effects like this can be properly tested, unlike our case where the late-time sampling is imperfect.

\subsection{Test of the standard model : radio behaviour, $\xi$ and low $n_0$}

The electron participation fraction, or $\xi$, can provide important clues regarding the nature of the burst and the circumburst medium. Our analyses in the previous section have shown that the afterglow emission can be equally well described by either of a top-hat or a Gaussian jet, with the latter having slightly higher energy requirements. Further, the best-fit including all data was found to have $\xi=1.0$ for both jet structures. A top-hat jet at $\xi=1.0$ was hence deemed the best model for the afterglow. However, the predicted (constant) CBM density was extremely low $<10^{-4}$ cm$^{-3}$ (or IGM-like), for both \texttt{afterglowpy} (Tables \ref{tab:pars_tophat}, \ref{tab:pars_gaussian_updated}) and analytical fits (Table \ref{tab:gerbil_results}). While this issue does not exist for the wind-like CBM, the latter provided a much worse fit to the data compared to the ISM, especially the X-rays, as in Figure \ref{fig:gerbillc}.

Further, upon ignoring the radio observations in the fit as in Figure \ref{fig:noradio}, the resulting best-fit density $n_0$ was much higher (than with the radio), $>0.01$ cm$^{-3}$ at $\xi=1.0$, which is still typically lower than expected ($\gtrsim0.1$ cm$^{-3}$, \citealt{ryden21}). All other parameters, including the energy budget, are either similar to that obtained previously and/or physically sound. The predicted radio emission in this case is almost 10 times higher than what is observed. For rare situations, low ISM densities $\sim10^{-3}$ cm$^{-3}$ may be associated with galactic halos that may harbor short GRBs \citep[e.g.,][]{debarros11}. With lower $\xi$, as in as low as $0.01$, the CBM density goes as high as $0.01$ cm$^{-3}$, which implies $n_0$ should range between $10^{-5}$ and $0.01$ cm$^{-3}$. Further, low CBM densities have been observed in many other GRBs previously (see e.g., \citealt{gompertz18}) -- and it has been argued that strong reverse-shock features in the light curves, in particular, necessitate a relatively low circumburst density \citep[e.g.,][]{laskar13,laskar16,alexander17}. It has been possible to find a set of best-fit parameters in \cite{gompertz18} that would have a much lower $\epsilon_B$ with an increased $n_0$, but situations like that were possible here only after making our priors smaller (e.g., increasing the lowest possible $n_0$). 

Even with such a low density (and a very narrow jet, $\theta_j \simeq 0.6^{\circ}$), the analytical fit, mostly based on \citet{granotsari02} and \citet{rhoads99}, furthermore fails to capture the full behavior of the afterglow. The radio and optical light curves are reproduced reasonably well, but the best fit places $\nu_c$ above 5~keV and, as a result, does not replicate the observed difference in the X-ray and optical pre-break slopes. As $\nu_c \propto n_0^{-1}$, this is a consequence of the low density required by the lower-frequency light curves. This is less of a problem in the \texttt{afterglowpy} fit (see Fig. \ref{fig:tophat_spec}).

It is hence unclear if the radio behaviour is not in accordance with the standard model, the reporting of very low ISM densities in GRBs incorporates a systematic error or if indeed a physical circumstance can produce ISM densities of $\lesssim10^{-4}$ cm$^{-3}$ in the environments where GRBs are produced. We refer to the work of \cite{kangas20} and \cite{kangas21}, which have shown that the radio afterglow behaviour is very peculiar for a large number of LGRBs, and even when the radio is fit well, the inferred physical parameters are often at the edge of a physically plausible prior. Our work adds to that by demonstrating a possible issue of the standard model dealing with the radio fluxes, which are overpredicted by an order of magnitude when a physically plausible CBM density is obtained or are fit well when the CBM density is $\lesssim10^{-4}$ cm$^{-3}$.

While the low density environment poses a challenge to the basic ISM model, we also point out that it may naturally arise in a ``wind bubble" scenario \citep{CMW75, Weaver+77}. In this scenario, the massive progenitor star of the GRB ejects a strong wind in its final stages of evolution. This wind clears a cavity around the progenitor, whose size is given by 
\begin{equation}
R_{bubble} \simeq 20~{\dot M}_{-8}^{1/5} v_{w,8}^{2/5} n_{0,1}^{-1/5} t_{\star,7}^{3/5}~{\rm pc}
\end{equation}
\citep[for a detailed discussion, see][]{PW06, PR24}. Here, $\dot M = 10^{-8} {\dot M}_{-8} M_\odot~{\rm yr}^{-1}$ is the mass ejection rate from the star, $v_w = 10^8~v_{w,8}~{\rm cm~s^{-1}}$ is the wind velocity, $t_\star = 10^7~t_{\star,7}$~yr is the lifetime of the wind ejection phase, and $n_0 = 10 n_{0,1}~{\rm cm^{-3}}$ sets the default ambient (ISM) density to $10~\mathrm{cm}^{-3}$. 
For these parameters, the cavity density is indeed very low, and given by 
\begin{equation}
 n_{cavity} \simeq 10^{-5}~{\dot M}_{-8}^{2/5} v_{w,8}^{-6/5} n_{0,1}^{3/5} t_{\star,7}^{-4/5},   
 \label{eq:n0}
\end{equation}
consistent with the values found here.

When the progenitor star explodes, the produced GRB blast wave then propagates into this low density cavity. For the fiducial parameters chosen, the total rest mass energy of cavity material, $E_{RM} = \dot M t_\star c^2 \approx 2 \times 10^{53}~{\dot M}_{-8}~t_{\star,7}$~erg is less than the isotropically equivalent energy released in the explosion producing the GRB. In this scenario, the GRB jet will cross the cavity while still relativistic, maintaining its original (coasting) Lorentz factor. The typical observed time scale in which the jet will cross the cavity is $T^{ob} \sim R_{bubble}/\Gamma^2 c \sim 3$~days, assuming $\Gamma = 100$; clearly, these values provide an order of magnitude estimation, and will vary based on the uncertain cavity size and GRB Lorentz factor.

Thus, the cavity scenario predicts an overall signal that is similar to that of an explosion into a low-density medium. The parameters chosen demonstrate that a cavity of several parsec size and very low density is plausible. A unique, a-chromatic signal may be observed when the blast wave crosses the cavity \citep{PW06, PR24}, but it may be difficult to detect \citep{NG07, VanEerten+09}.

\section{Conclusions}
\begin{enumerate}
    \item GRB 220101A is one of the most energetic bursts ever detected, with $E_{\gamma,\,\rm iso}\sim3.6\times10^{54}$ erg at a redshift of $z=4.618$. The afterglow light curve shows a distinct steepening at $\gtrsim8.6\pm0.1$ days from radio through X-rays, with a post-break slope close to $\sim3.0\pm0.1$, which is $>5\sigma$ steeper than the slope of $p$ ($2.50$ from analytical spectral fitting) expected from simple lateral expansion theory of a top-hat jet. This hints at possibly more complex effects in play that include hydrodynamics and geometric effects (e.g., \citealt{zhang09}).
    \item We fit model afterglow light curves for a top-hat and Gaussian jet using \texttt{afterglowpy} \citep{ryan20} to better understand the physics of the GRB engine. We found that the case with the highest electron participation fraction $\xi=1.0$ described the light curve and spectral behaviour the best, with physically feasible parameters enlisted in Tables 2-4, except the case of the CBM density, which was found to be $<10^{-4}$ cm$^{-3}$ for the $\xi=1.0$ case. Removal of the radio data from the fit significantly increased the density -- by two orders of magnitude -- with an overprediction of the radio in that case. 
    
    \item Similar behaviour is common in other LAT bursts \citep{gompertz18}, including GRB 160625B \citep{kangas20}, where the densities were similarly low. It is either possible that GRBs commonly inhabit large cavities in the IGM or the radio behaviour is peculiar and we do not have an explanation for the same using the standard model. However, while we are modeling a time-dependent source, the microphysics parameters are assumed to be static. It is hence worth exploring whether introducing a time dependency in some parameters could ameliorate the issues encountered in modeling radio afterglows.

    \item A possible self-consistent explanation for such low ISM densities lies in the possible evacuation of the circumburst medium by strong winds launched by the GRB progenitor in the final stages of its evolution. The GRB jet is then expected to propagate into this low density cavity. Appropriate sets of physical parameters (like mass ejection rate, wind velocity or lifetime of the wind ejection phase) can in principle lead to ISM densities that are $\sim10^{-5}$ cm$^{-3}$ (Equation \ref{eq:n0}). However, due to the parameter uncertainties involved (as discussed in the text), we cannot say for sure if the low densities are indeed due to progenitor winds or are a problem of the standard model.
    
    \item It is possible we are seeing peculiar radio behaviour in the afterglow of GRB 220101A, as mentioned above. This ties in with the general absence of complete understanding of the radio emission from GRBs (see e.g., \citealt{rhodes24}). Furthermore, radio-loud and radio-quiet GRBs, as discussed in a number of previous studies \citep[e.g.,][]{hancock13, gompertz18, lloyd19, lloyd22}, have been suggested to originate from different GRB progenitors, where the former possibly originates from a binary merger. Differences in the nature of progenitors necessarily include differences in the physical properties of these GRBs, e.g., the isotropic energies, spectral shapes, or the circumburst density profile. A larger CBM (ISM-like) density requires a much brighter-than-observed radio emission in the case of GRB220101A according to the standard fireball model. In contrast, if the radio flux level is predicted correctly by the standard model, the CBM densities become very low. It is possible this is related to a problem in the standard model, or that the progenitors of GRBs can accommodate wide differences in the properties of the circumburst medium profile \citep{fryer98,zhang01,lloyd19}. 
\end{enumerate}

\begin{acknowledgements}

ARC acknowledges the support of the Department of Atomic Energy, Government of India, under the project 12-R\&D-TFR-5.02-0700 and several pieces of advice from Geoffrey Ryan. TK acknowledges support from the Research Council of Finland project 360274. This paper makes use of the following ALMA data: ADS/JAO.ALMA\#2021.1.00658.T. ALMA is a partnership of ESO (representing its member states), NSF (USA) and NINS (Japan), together with NRC (Canada), NSTC and ASIAA (Taiwan), and KASI (Republic of Korea), in cooperation with the Republic of Chile. The Joint ALMA Observatory is operated by ESO, AUI/NRAO and NAOJ. The National Radio Astronomy Observatory and Green Bank Observatory are facilities of the U.S. National Science Foundation operated under cooperative agreement by Associated Universities, Inc. \\

\end{acknowledgements}

\appendix
\iffalse

\section{Corner Plots}

\begin{figure*}
\includegraphics[width=\textwidth]{for_paper_tophat_xi0.1_corner.pdf} 
    \caption{}
    \label{fig:xi0.1}
\end{figure*}

\begin{figure*}
\includegraphics[width=\textwidth]{for_paper_tophat_x1.0_corner.pdf} 
    \caption{}
    \label{fig:xi1.0}
\end{figure*}

\fi

\section{Comparison with the results of Zhu et al.}
\label{app:zhu}

\begin{figure}
    \centering
    \includegraphics[width=0.49\linewidth]{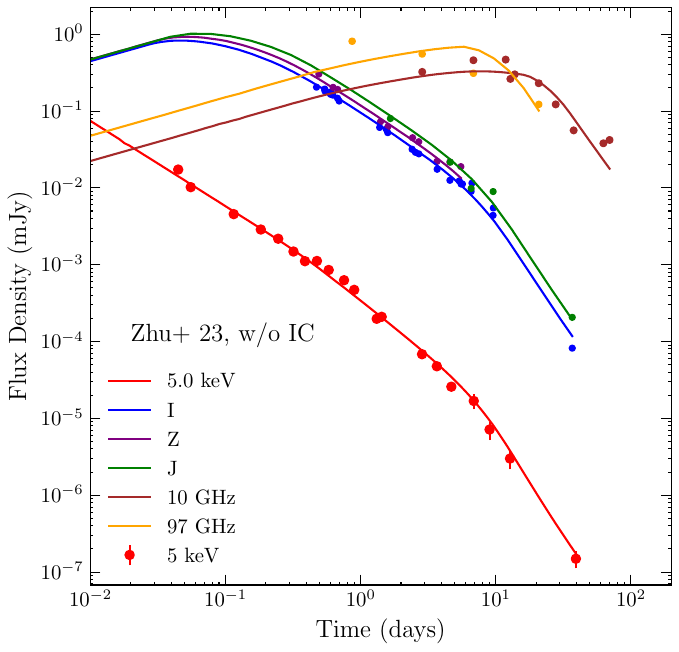}
    \caption{Afterglow light curve of GRB 220101A with the initial parameters taken from \cite{zhu23} and run for 300 walkers and 2000 steps. The best-fitting \texttt{afterglowpy} model is shown in solid lines when inverse Compton losses are ignored.}
    \label{fig:zhu}
\end{figure}

\texttt{Afterglowpy} (with $\xi=1.0$) fits to the same afterglow by \cite{zhu23}, where the only difference was the absence of the final Chandra X-ray data point at $\sim40$ days, show markedly different set of parameters, especially the inferred ISM number density, which was $\gtrsim 10^4$ times the one obtained using our fits. With the intention of examining this discrepancy, we fit our data starting from the \cite{zhu23} best-fit parameters for two spectral types inside \texttt{afterglowpy}, with and without inverse Compton losses. For the case where IC losses are ignored, the spectrum fits well with the \cite{zhu23} parameters ($\chi^2$/DOF $\simeq 2$) and the radio is well reproduced (Figure \ref{fig:zhu}) with the number density ``physical". In contrast, for the more physically relevant scenario that includes IC losses, the radio fit is poor (and $\chi^2$/DOF $\simeq 9$) for a low walker-step pair (300, 2000), with higher density values still, implying the \cite{zhu23} parameters did not form the global minima. For a longer and larger run, the simulation converges to the values we have obtained in the main sections of the paper, which is expected.

Table \ref{tab:pars_zhu_comp} shows the best-fit parameters for our run with IC losses, the parameters without IC losses, and the parameters from \cite{zhu23} in three separate columns. It is possible the question therefore partly boils down to inclusion of IC losses in the fits. However, from a purely physical standpoint, it would be unwise to ignore IC losses, howsoever physical a parameter space one might obtain by ignoring IC losses. We compared our results with that of \cite{zhu23} to only throw light on the parameter discrepancies for the same source.

\begin{table}
    \centering
    \footnotesize
    \begin{tabularx}{0.5\textwidth}{c|ccc}
    \hline
    Parameter & With IC & Without IC & \cite{zhu23} \\
    \hline
    $\log (E_{\rm K, iso} / \text{erg})$ & $56.17^{+0.07}_{-0.05}$ & $55.02^{+0.07}_{-0.07}$ & $54.57^{+0.03}_{-0.03}$ \\
    $p$ & $2.16\pm 0.02$ & $2.14^{+0.02}_{-0.02}$ & $2.44^{+0.01}_{-0.01}$ \\
    $\log (\theta_{\rm obs} / \text{rad})$ & $-2.45\pm 0.04$ & $-1.73\pm0.05$ & $\theta_{\rm obs}=0$ \\
    $\log (\theta_j / \text{rad})$ & $-2.19\pm0.03$ & $-1.46^{+0.04}_{-0.06}$ & $-1.22^{+0.07}_{-0.08}$ \\
    $\log (n_0 / \text{cm}^{-3})$ & $-5.44^{+0.25}_{-0.33}$ & $-0.88^{+0.27}_{-0.31}$ & $-1.01^{+0.33}_{-0.39}$ \\
    $\log \epsilon_e$ & $-1.38^{+0.11}_{-0.09}$ & $-0.48^{+0.08}_{-0.07}$ & $-0.51^{+0.02}_{-0.02}$ \\
    $\log \epsilon_B$ & $-1.84^{+0.34}_{-0.31}$ & $-4.03^{+0.41}_{-0.16}$ & $-4.13^{+0.17}_{-0.37}$ \\
    $A_B$ & $0.13^{+0.04}_{-0.02}$ & $0.15^{+0.03}_{-0.03}$ & \\
    \hline
    $E_{\rm K}$ (erg) & $3.1\times10^{51}$ & $6.2\times10^{51}$ & $6.3\times10^{51}$ \\ 
    $\chi^2$/DOF & 2.1 & 2.0 & \\ 
    \hline
    \end{tabularx}
    \caption{Best-fit parameters for the Tophat jet for different values of $\xi$.}
    \label{tab:pars_zhu_comp}
\end{table}

\section{The afterglow of GRB 160625B}
\label{app:160625b}

Similar peculiar ISM density behaviour in GRB 160625B had been observed as well. Here we ran \texttt{afterglowpy} fits (with 300 walkers and 3000 steps) to the afterglow of the source for $\xi=1.0$, with and without the radio data.

\begin{figure}
    \centering
    \vbox
    {
    \includegraphics[width=0.49\linewidth]{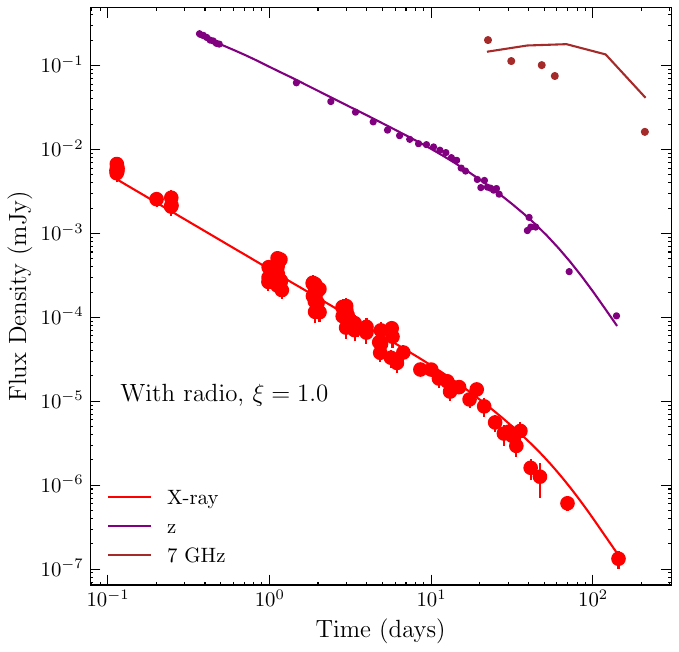}
    \includegraphics[width=0.49\linewidth]{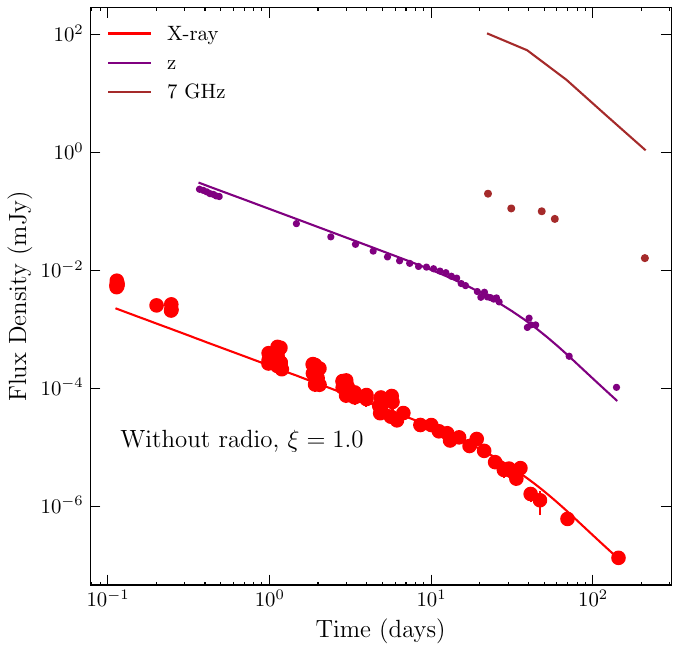}
    }
    \caption{Afterglow light curve of GRB 160625B with the best-fitting \texttt{afterglowpy} model including and excluding the radio data in the left and right panels respectively. Exclusion of the radio data shows strong overprediction ($\sim$ 100 times) of the observed radio flux densities.}
    \label{fig:160625B}
\end{figure}

Figure \ref{fig:160625B} shows the difference in fits for $\xi=1.0$ when the radio is considered and when it is not. Parameter results have been tabulated below in Table \ref{tab:pars_radio_1606}. The issues are very similar to GRB220101A: extremely low $n_0$ when the radio is included and a ``physical" $n_0$ when excluded with $\sim100$ times overprediction of the radio flux densities.

\begin{table}
    \centering
    \footnotesize
    \begin{tabularx}{0.4\textwidth}{c|cc}
    \hline
    Parameter & Without Radio & With Radio \\
    \hline
    $\log (E_{\rm K, iso} / \text{erg})$ & $55.61^{+0.27}_{-0.28}$ & $55.10^{+0.10}_{-0.10}$ \\
    $p$ & $2.01^{+0.27}_{-0.01}$ & $2.28^{+0.04}_{-0.02}$ \\
    $\log (\theta_{\rm obs} / \text{rad})$ & $-1.66^{+0.15}_{-0.15}$ & $-2.02^{+0.07}_{-0.06}$ \\
    $\log (\theta_j / \text{rad})$ & $-1.06^{+0.09}_{-0.36}$ & $-1.46^{+0.06}_{-0.06}$ \\
    $\log (n_0 / \text{cm}^{-3})$ & $-0.95^{+0.47}_{-3.69}$ & $-4.98^{+0.09}_{-0.02}$ \\
    $\log \epsilon_e$ & $-0.65^{+0.18}_{-0.38}$ & $-0.54^{+0.04}_{-0.04}$ \\
    $\log \epsilon_B$ & $-2.46^{+0.46}_{-0.46}$ & $-3.10^{+0.15}_{-0.15}$ \\
    $A_V$ & $1.00^{+0.00}_{-0.91}$ & $0.07^{+0.01}_{-0.01}$ \\
    \hline
    $E_{\rm K}$ (erg) & $1.5\times10^{53}$ & $7.5\times10^{51}$ \\
    $\chi^2$/DOF & 7.9 & 2.1 \\
    \hline
    \end{tabularx}
    \caption{Best-fit parameters for models with and without radio data for GRB160625B.}
    \label{tab:pars_radio_1606}
\end{table}

\bibliographystyle{aasjournal}
\bibliography{version1}

@ARTICLE{Burns23,
       author = {{Burns}, Eric and {Svinkin}, Dmitry and {Fenimore}, Edward and {Kann}, D. Alexander and {Ag{\"u}{\'\i} Fern{\'a}ndez}, Jos{\'e} Feliciano and {Frederiks}, Dmitry and {Hamburg}, Rachel and {Lesage}, Stephen and {Temiraev}, Yuri and {Tsvetkova}, Anastasia and {Bissaldi}, Elisabetta and {Briggs}, Michael S. and {Dalessi}, Sarah and {Dunwoody}, Rachel and {Fletcher}, Cori and {Goldstein}, Adam and {Hui}, C. Michelle and {Hristov}, Boyan A. and {Kocevski}, Daniel and {Lysenko}, Alexandra L. and {Mailyan}, Bagrat and {Mangan}, Joseph and {McBreen}, Sheila and {Racusin}, Judith and {Ridnaia}, Anna and {Roberts}, Oliver J. and {Ulanov}, Mikhail and {Veres}, Peter and {Wilson-Hodge}, Colleen A. and {Wood}, Joshua},
        title = "{GRB 221009A: The Boat}",
      journal = {\apjl},
         year = 2023,
        month = mar,
       volume = {946},
       number = {1},
          eid = {L31},
        pages = {L31},
          doi = {10.3847/2041-8213/acc39c},
archivePrefix = {arXiv},
       eprint = {2302.14037},
 primaryClass = {astro-ph.HE},
       adsurl = {https://ui.adsabs.harvard.edu/abs/2023ApJ...946L..31B}
}

@ARTICLE{PW06,
   author = {{Pe'er}, A. and {Wijers}, R.~A.~M.~J.},
    title = "{The Signature of a Wind Reverse Shock in Gamma-Ray Burst Afterglows}",
  journal = {\apj},
   eprint = {arXiv:astro-ph/0511508},
     year = 2006,
    month = jun,
   volume = 643,
    pages = {1036-1046},
      doi = {10.1086/500969},
   adsurl = {http://adsabs.harvard.edu/abs/2006ApJ...643.1036P}
}

@ARTICLE{PR24,
       author = {{Pe'er}, Asaf and {Ryde}, Felix},
        title = "{Gamma-ray burst interaction with the circumburst medium: The CBM phase of GRBs}",
      journal = {arXiv e-prints},
         year = 2024,
        month = jun,
          eid = {arXiv:2406.03841},
        pages = {arXiv:2406.03841},
          doi = {10.48550/arXiv.2406.03841},
archivePrefix = {arXiv},
       eprint = {2406.03841},
 primaryClass = {astro-ph.HE},
       adsurl = {https://ui.adsabs.harvard.edu/abs/2024arXiv240603841P}
}

@ARTICLE{NG07,
       author = {{Nakar}, Ehud and {Granot}, Jonathan},
        title = "{Smooth light curves from a bumpy ride: relativistic blast wave encounters a density jump}",
      journal = {\mnras},
         year = 2007,
        month = oct,
       volume = {380},
       number = {4},
        pages = {1744-1760},
          doi = {10.1111/j.1365-2966.2007.12245.x},
archivePrefix = {arXiv},
       eprint = {astro-ph/0606011},
 primaryClass = {astro-ph},
       adsurl = {https://ui.adsabs.harvard.edu/abs/2007MNRAS.380.1744N}
}

@ARTICLE{VanEerten+09,
   author = {{van Eerten}, H.~J. and {Meliani}, Z. and {Wijers}, R.~A.~M.~J. and 
	{Keppens}, R.},
    title = "{No visible optical variability from a relativistic blast wave encountering a wind termination shock}",
  journal = {\mnras},
archivePrefix = "arXiv",
   eprint = {0906.3629},
 primaryClass = "astro-ph.HE",
     year = 2009,
    month = sep,
   volume = 398,
    pages = {L63-L67},
      doi = {10.1111/j.1745-3933.2009.00711.x},
   adsurl = {http://adsabs.harvard.edu/abs/2009MNRAS.398L..63V}
}

@ARTICLE{CMW75,
       author = {{Castor}, J. and {McCray}, R. and {Weaver}, R.},
        title = "{Interstellar bubbles.}",
      journal = {\apjl},
         year = 1975,
        month = sep,
       volume = {200},
        pages = {L107-L110},
          doi = {10.1086/181908},
       adsurl = {https://ui.adsabs.harvard.edu/abs/1975ApJ...200L.107C}
}

@ARTICLE{Weaver+77,
   author = {{Weaver}, R. and {McCray}, R. and {Castor}, J. and {Shapiro}, P. and 
	{Moore}, R.},
    title = "{Interstellar bubbles. II - Structure and evolution}",
  journal = {\apj},
     year = 1977,
    month = dec,
   volume = 218,
    pages = {377-395},
      doi = {10.1086/155692},
   adsurl = {http://adsabs.harvard.edu/abs/1977ApJ...218..377W}
}

@ARTICLE{kumarzhang15,
       author = {{Kumar}, Pawan and {Zhang}, Bing},
        title = "{The physics of gamma-ray bursts \& relativistic jets}",
      journal = {\physrep},
         year = 2015,
        month = feb,
       volume = {561},
        pages = {1-109},
          doi = {10.1016/j.physrep.2014.09.008},
archivePrefix = {arXiv},
       eprint = {1410.0679},
 primaryClass = {astro-ph.HE},
       adsurl = {https://ui.adsabs.harvard.edu/abs/2015PhR...561....1K}
}

@BOOK{ryden21,
       author = {{Ryden}, Barbara Sue and {Pogge}, Richard W.},
        title = "{Interstellar and intergalactic medium}",
         year = 2021,
    publisher = {Cambridge University Press},
          doi = {10.1017/9781108781596},
       adsurl = {https://ui.adsabs.harvard.edu/abs/2021iim..book.....R}
}

@ARTICLE{hancock13,
       author = {{Hancock}, P.~J. and {Gaensler}, B.~M. and {Murphy}, T.},
        title = "{Two Populations of Gamma-Ray Burst Radio Afterglows}",
      journal = {\apj},
         year = 2013,
        month = oct,
       volume = {776},
       number = {2},
          eid = {106},
        pages = {106},
          doi = {10.1088/0004-637X/776/2/106},
archivePrefix = {arXiv},
       eprint = {1308.4766},
 primaryClass = {astro-ph.HE},
       adsurl = {https://ui.adsabs.harvard.edu/abs/2013ApJ...776..106H}
}

@ARTICLE{laskar13,
       author = {{Laskar}, T. and {Berger}, E. and {Zauderer}, B.~A. and {Margutti}, R. and {Soderberg}, A.~M. and {Chakraborti}, S. and {Lunnan}, R. and {Chornock}, R. and {Chandra}, P. and {Ray}, A.},
        title = "{A Reverse Shock in GRB 130427A}",
      journal = {\apj},
         year = 2013,
        month = oct,
       volume = {776},
       number = {2},
          eid = {119},
        pages = {119},
          doi = {10.1088/0004-637X/776/2/119},
archivePrefix = {arXiv},
       eprint = {1305.2453},
 primaryClass = {astro-ph.HE},
       adsurl = {https://ui.adsabs.harvard.edu/abs/2013ApJ...776..119L}
}

@ARTICLE{laskar16,
       author = {{Laskar}, Tanmoy and {Alexander}, Kate D. and {Berger}, Edo and {Fong}, Wen-fai and {Margutti}, Raffaella and {Shivvers}, Isaac and {Williams}, Peter K.~G. and {Kopa{\v{c}}}, Drejc and {Kobayashi}, Shiho and {Mundell}, Carole and {Gomboc}, Andreja and {Zheng}, WeiKang and {Menten}, Karl M. and {Graham}, Melissa L. and {Filippenko}, Alexei V.},
        title = "{A Reverse Shock in GRB 160509A}",
      journal = {\apj},
         year = 2016,
        month = dec,
       volume = {833},
       number = {1},
          eid = {88},
        pages = {88},
          doi = {10.3847/1538-4357/833/1/88},
archivePrefix = {arXiv},
       eprint = {1606.08873},
 primaryClass = {astro-ph.HE},
       adsurl = {https://ui.adsabs.harvard.edu/abs/2016ApJ...833...88L}
}

@ARTICLE{lloyd22,
       author = {{Lloyd-Ronning}, Nicole},
        title = "{Radio-loud versus Radio-quiet Gamma-Ray Bursts: The Role of Binary Progenitors}",
      journal = {\apj},
         year = 2022,
        month = apr,
       volume = {928},
       number = {2},
          eid = {104},
        pages = {104},
          doi = {10.3847/1538-4357/ac54b3},
archivePrefix = {arXiv},
       eprint = {2109.14122},
 primaryClass = {astro-ph.HE},
       adsurl = {https://ui.adsabs.harvard.edu/abs/2022ApJ...928..104L}
}

@ARTICLE{lloyd19,
       author = {{Lloyd-Ronning}, Nicole M. and {Gompertz}, Ben and {Pe'er}, Asaf and {Dainotti}, Maria and {Fruchter}, Andy},
        title = "{A Comparison between Radio Loud and Quiet Gamma-Ray Bursts, and Evidence for a Potential Correlation between Intrinsic Duration and Redshift in the Radio Loud Population}",
      journal = {\apj},
         year = 2019,
        month = jan,
       volume = {871},
       number = {1},
          eid = {118},
        pages = {118},
          doi = {10.3847/1538-4357/aaf6ac},
       adsurl = {https://ui.adsabs.harvard.edu/abs/2019ApJ...871..118L}
}

@ARTICLE{fryer98,
       author = {{Fryer}, C.~L. and {Woosley}, S.~E.},
        title = "{Helium Star/Black Hole Mergers: A New Gamma-Ray Burst Model}",
      journal = {\apjl},
         year = 1998,
        month = jul,
       volume = {502},
       number = {1},
        pages = {L9-L12},
          doi = {10.1086/311493},
archivePrefix = {arXiv},
       eprint = {astro-ph/9804167},
 primaryClass = {astro-ph},
       adsurl = {https://ui.adsabs.harvard.edu/abs/1998ApJ...502L...9F}
}

@ARTICLE{zhang01,
       author = {{Zhang}, Weiqun and {Fryer}, Chris L.},
        title = "{The Merger of a Helium Star and a Black Hole: Gamma-Ray Bursts}",
      journal = {\apj},
         year = 2001,
        month = mar,
       volume = {550},
       number = {1},
        pages = {357-367},
          doi = {10.1086/319734},
archivePrefix = {arXiv},
       eprint = {astro-ph/0011236},
 primaryClass = {astro-ph},
       adsurl = {https://ui.adsabs.harvard.edu/abs/2001ApJ...550..357Z}
}

@ARTICLE{sari98,
       author = {{Sari}, Re'em and {Piran}, Tsvi and {Narayan}, Ramesh},
        title = "{Spectra and Light Curves of Gamma-Ray Burst Afterglows}",
      journal = {\apjl},
         year = 1998,
        month = apr,
       volume = {497},
       number = {1},
        pages = {L17-L20},
          doi = {10.1086/311269},
archivePrefix = {arXiv},
       eprint = {astro-ph/9712005},
 primaryClass = {astro-ph},
       adsurl = {https://ui.adsabs.harvard.edu/abs/1998ApJ...497L..17S}
}

@ARTICLE{alexander17,
       author = {{Alexander}, K.~D. and {Laskar}, T. and {Berger}, E. and {Guidorzi}, C. and {Dichiara}, S. and {Fong}, W. and {Gomboc}, A. and {Kobayashi}, S. and {Kopac}, D. and {Mundell}, C.~G. and {Tanvir}, N.~R. and {Williams}, P.~K.~G.},
        title = "{A Reverse Shock and Unusual Radio Properties in GRB 160625B}",
      journal = {\apj},
         year = 2017,
        month = oct,
       volume = {848},
       number = {1},
          eid = {69},
        pages = {69},
          doi = {10.3847/1538-4357/aa8a76},
archivePrefix = {arXiv},
       eprint = {1705.08455},
 primaryClass = {astro-ph.HE},
       adsurl = {https://ui.adsabs.harvard.edu/abs/2017ApJ...848...69A}
}

@ARTICLE{tsvetkova22gcn,
       author = {{Tsvetkova}, A. and {Frederiks}, D. and {Lysenko}, A. and {Ridnaia}, A. and {Svinkin}, D. and {Ulanov}, M. and {Cline}, T. and {Konus-Wind Team}},
        title = "{Konus-Wind detection of GRB 220101A}",
      journal = {GRB Coordinates Network},
         year = 2022,
        month = jan,
       volume = {31433},
        pages = {1},
       adsurl = {https://ui.adsabs.harvard.edu/abs/2022GCN.31433....1T}
}

@ARTICLE{daly88,
       author = {{Daly}, Ruth A. and {Marscher}, Alan P.},
        title = "{The Gasdynamics of Compact Relativistic Jets}",
      journal = {\apj},
         year = 1988,
        month = nov,
       volume = {334},
        pages = {539},
          doi = {10.1086/166858},
       adsurl = {https://ui.adsabs.harvard.edu/abs/1988ApJ...334..539D}
}

@ARTICLE{rhodes24,
       author = {{Rhodes}, L. and {van der Horst}, A.~J. and {Bright}, J.~S. and {Leung}, J.~K. and {Anderson}, G.~E. and {Fender}, R. and {Ag{\"u}{\'\i} Fern{\'a}ndez}, J.~F. and {Bremer}, M. and {Chandra}, P. and {Dobie}, D. and {Farah}, W. and {Giarratana}, S. and {Gourdji}, K. and {Green}, D.~A. and {Lenc}, E. and {Micha{\l}owski}, M.~J. and {Murphy}, T. and {Nayana}, A.~J. and {Pollak}, A.~W. and {Rowlinson}, A. and {Schussler}, F. and {Siemion}, A. and {Starling}, R.~L.~C. and {Scott}, P. and {Th{\"o}ne}, C.~C. and {Titterington}, D. and {de Ugarte Postigo}, A.},
        title = "{Rocking the BOAT: the ups and downs of the long-term radio light curve for GRB 221009A}",
      journal = {\mnras},
         year = 2024,
        month = oct,
       volume = {533},
       number = {4},
        pages = {4435-4449},
          doi = {10.1093/mnras/stae2050},
archivePrefix = {arXiv},
       eprint = {2408.16637},
 primaryClass = {astro-ph.HE},
       adsurl = {https://ui.adsabs.harvard.edu/abs/2024MNRAS.533.4435R}
}

@ARTICLE{markwardt22gcn,
       author = {{Markwardt}, C.~B. and {Barthelmy}, S.~D. and {Krimm}, H.~A. and {Laha}, S. and {Lien}, A.~Y. and {Palmer}, D.~M. and {Parsotan}, T. and {Sakamoto}, T. and {Stamatikos}, M.},
        title = "{GRB 220101A: Swift-BAT refined analysis}",
      journal = {GRB Coordinates Network},
         year = 2022,
        month = jan,
       volume = {31369},
        pages = {1},
       adsurl = {https://ui.adsabs.harvard.edu/abs/2022GCN.31369....1M}
}

@ARTICLE{zhu23,
       author = {{Zhu}, Zi-Pei and {Lei}, Wei-Hua and {Malesani}, Daniele B. and {Fu}, Shao-Yu and {Liu}, Dong-Jie and {Xu}, Dong and {D'Avanzo}, Paolo and {Ag{\"u}{\'\i} Fern{\'a}ndez}, Jos{\'e} Feliciano and {Fynbo}, Johan P.~U. and {Gao}, Xing and {Nicuesa Guelbenzu}, Ana and {Jiang}, Shuai-Qing and {Kann}, David Alexander and {Klose}, Sylvio and {Liu}, Jin-Zhong and {Liu}, Xing and {De Pasquale}, Massimiliano and {de Ugarte Postigo}, Antonio and {Stecklum}, Bringfried and {Th{\"o}ne}, Christina and {Markku Viuho}, Joonas Kari and {Zhu}, Yi-Nan and {Li}, Jin-Da and {Gao}, He and {Lu}, Tian-Hua and {Xiao}, Shuo and {Zou}, Yuan-Chuan and {Xin}, Li-Ping and {Wei}, Jian-Yan},
        title = "{Optical and Near-infrared Observations of the Distant but Bright ``New Year's Burst'' GRB 220101A}",
      journal = {\apj},
         year = 2023,
        month = dec,
       volume = {959},
       number = {2},
          eid = {118},
        pages = {118},
          doi = {10.3847/1538-4357/ad05c8},
archivePrefix = {arXiv},
       eprint = {2303.09982},
 primaryClass = {astro-ph.HE},
       adsurl = {https://ui.adsabs.harvard.edu/abs/2023ApJ...959..118Z}
}

@ARTICLE{fynbo22gcn,
       author = {{Fynbo}, J.~P.~U. and {de Ugarte Postigo}, A. and {Xu}, D. and {Malesani}, D.~B. and {Milvang-Jensen}, B. and {Viuho}, J.},
        title = "{GRB 220101A: NOT redshift confirmation}",
      journal = {GRB Coordinates Network},
         year = 2022,
        month = jan,
       volume = {31359},
        pages = {1},
       adsurl = {https://ui.adsabs.harvard.edu/abs/2022GCN.31359....1F}
}

@ARTICLE{tohuvavohu22gcn,
       author = {{Tohuvavohu}, A. and {Gropp}, J.~D. and {Kennea}, J.~A. and {Lien}, A.~Y. and {Palmer}, D.~M. and {Parsotan}, T.~M. and {Sbarufatti}, B. and {Siegel}, M.~H. and {Neil Gehrels Swift Observatory Team}},
        title = "{GRB 220101A: Swift detection of a burst with a bright optical counterpart}",
      journal = {GRB Coordinates Network},
         year = 2022,
        month = jan,
       volume = {31347},
        pages = {1},
       adsurl = {https://ui.adsabs.harvard.edu/abs/2022GCN.31347....1T}
}

@ARTICLE{perley22gcn2,
       author = {{Perley}, D.~A.},
        title = "{GRB 220101A: Additional Liverpool telescope photometry}",
      journal = {GRB Coordinates Network},
         year = 2022,
        month = jan,
       volume = {31425},
        pages = {1},
       adsurl = {https://ui.adsabs.harvard.edu/abs/2022GCN.31425....1P}
}

@ARTICLE{guelbenzu22gcn,
       author = {{Nicuesa Guelbenzu}, A. and {Melnikov}, S. and {Klose}, S. and {Stecklum}, B. and {Ludwig}, F.},
        title = "{GRB 220101A: Tautenburg observations}",
      journal = {GRB Coordinates Network},
         year = 2022,
        month = jan,
       volume = {31401},
        pages = {1},
       adsurl = {https://ui.adsabs.harvard.edu/abs/2022GCN.31401....1N}
}

@ARTICLE{davanzo22gcn2,
       author = {{D'Avanzo}, P. and {Melandri}, A. and {Fugazza}, D. and {Campana}, S. and {Malesani}, D.~B. and {D'Elia}, V. and {De Pasquale}, M. and {Palazzi}, E. and {Piranomonte}, S. and {Rossi}, A. and {Tagliaferri}, G. and {Lorenzi} and {V.} and {Carosati}, D. and {Giacobbe}, Paolo and {CIBO Collaboration}},
        title = "{GRB 220101A: further TNG NIR observations}",
      journal = {GRB Coordinates Network},
         year = 2022,
        month = jan,
       volume = {31395},
        pages = {1},
       adsurl = {https://ui.adsabs.harvard.edu/abs/2022GCN.31395....1D}
}

@ARTICLE{davanzo22gcn,
       author = {{D'Avanzo}, P. and {Melandri}, A. and {Covino}, S. and {D'Elia}, V. and {De Pasquale}, M. and {Malesani}, D.~B. and {Piranomonte}, S. and {Lorenzi}, V. and {Padilla}, C. and {Giacobbe}, Paolo and {CIBO Collaboration}},
        title = "{GRB 220101A: TNG NIR afterglow detection}",
      journal = {GRB Coordinates Network},
         year = 2022,
        month = jan,
       volume = {31373},
        pages = {1},
       adsurl = {https://ui.adsabs.harvard.edu/abs/2022GCN.31373....1D}
}

@ARTICLE{vinko22gcn,
       author = {{Vinko}, J. and {Pal}, A. and {Kriskovics}, L. and {Szakats}, R. and {Vida}, K.},
        title = "{GRB220101A: optical afterglow detection from Konkoly Observatory}",
      journal = {GRB Coordinates Network},
         year = 2022,
        month = jan,
       volume = {31361},
        pages = {1},
       adsurl = {https://ui.adsabs.harvard.edu/abs/2022GCN.31361....1V}
}

@ARTICLE{postigo22gcn,
       author = {{de Ugarte Postigo}, A. and {Kann}, D.~A. and {Thoene}, C.~C. and {Blazek}, M. and {Agui Fernandez}, J.~F. and {Martin}, P. and {Hermelo}, I.},
        title = "{GRB 220101A: CAHA 2.2m/CAFOS detection}",
      journal = {GRB Coordinates Network},
         year = 2022,
        month = jan,
       volume = {31358},
        pages = {1},
       adsurl = {https://ui.adsabs.harvard.edu/abs/2022GCN.31358....1D}
}

@ARTICLE{perley22gcn,
       author = {{Perley}, D.~A.},
        title = "{GRB 220101A: Liverpool telescope imaging of a high-redshift afterglow}",
      journal = {GRB Coordinates Network},
         year = 2022,
        month = jan,
       volume = {31357},
        pages = {1},
       adsurl = {https://ui.adsabs.harvard.edu/abs/2022GCN.31357....1P}
}

@ARTICLE{fu22gcn,
       author = {{Fu}, S.~Y. and {Zhu}, Z.~P. and {Xu}, D. and {Liu}, X. and {Jiang}, S.~Q.},
        title = "{GRB 220101A: Xinglong-2.16m photometry and spectroscopy}",
      journal = {GRB Coordinates Network},
         year = 2022,
        month = jan,
       volume = {31353},
        pages = {1},
       adsurl = {https://ui.adsabs.harvard.edu/abs/2022GCN.31353....1F}
}

@ARTICLE{evans07,
   author = {{Evans}, P.~A. and {Beardmore}, A.~P. and {Page}, K.~L. and 
	{Tyler}, L.~G. and {Osborne}, J.~P. and {Goad}, M.~R. and {O'Brien}, P.~T. and 
	{Vetere}, L. and {Racusin}, J. and {Morris}, D. and {Burrows}, D.~N. and 
	{Capalbi}, M. and {Perri}, M. and {Gehrels}, N. and {Romano}, P.
	},
    title = "{An online repository of Swift/XRT light curves of {$\gamma$}-ray bursts}",
  journal = {\aap},
archivePrefix = "arXiv",
   eprint = {0704.0128},
     year = 2007,
    month = jul,
   volume = 469,
    pages = {379-385},
      doi = {10.1051/0004-6361:20077530},
   adsurl = {http://adsabs.harvard.edu/abs/2007A%26A...469..379E}
}

@ARTICLE{evans09,
   author = {{Evans}, P.~A. and {Beardmore}, A.~P. and {Page}, K.~L. and 
	{Osborne}, J.~P. and {O'Brien}, P.~T. and {Willingale}, R. and 
	{Starling}, R.~L.~C. and {Burrows}, D.~N. and {Godet}, O. and 
	{Vetere}, L. and {Racusin}, J. and {Goad}, M.~R. and {Wiersema}, K. and 
	{Angelini}, L. and {Capalbi}, M. and {Chincarini}, G. and {Gehrels}, N. and 
	{Kennea}, J.~A. and {Margutti}, R. and {Morris}, D.~C. and {Mountford}, C.~J. and 
	{Pagani}, C. and {Perri}, M. and {Romano}, P. and {Tanvir}, N.
	},
    title = "{Methods and results of an automatic analysis of a complete sample of Swift-XRT observations of GRBs}",
  journal = {\mnras},
archivePrefix = "arXiv",
   eprint = {0812.3662},
     year = 2009,
    month = aug,
   volume = 397,
    pages = {1177-1201},
      doi = {10.1111/j.1365-2966.2009.14913.x},
   adsurl = {http://adsabs.harvard.edu/abs/2009MNRAS.397.1177E}
}

@ARTICLE{kangas21,
       author = {{Kangas}, Tuomas and {Fruchter}, Andrew S.},
        title = "{The Late-time Radio Behavior of Gamma-ray Burst Afterglows: Testing the Standard Model}",
      journal = {\apj},
         year = 2021,
        month = apr,
       volume = {911},
       number = {1},
          eid = {14},
        pages = {14},
          doi = {10.3847/1538-4357/abe76b},
archivePrefix = {arXiv},
       eprint = {1911.01938},
 primaryClass = {astro-ph.HE},
       adsurl = {https://ui.adsabs.harvard.edu/abs/2021ApJ...911...14K}
}

@ARTICLE{granotsari02,
   author = {{Granot}, J. and {Sari}, R.},
    title = "{The Shape of Spectral Breaks in Gamma-Ray Burst Afterglows}",
  journal = {\apj},
   eprint = {astro-ph/0108027},
     year = 2002,
    month = apr,
   volume = 568,
    pages = {820-829},
      doi = {10.1086/338966},
   adsurl = {http://adsabs.harvard.edu/abs/2002ApJ...568..820G}
}

@ARTICLE{foremanmackey13,
       author = {{Foreman-Mackey}, Daniel and {Hogg}, David W. and {Lang}, Dustin and {Goodman}, Jonathan},
        title = "{emcee: The MCMC Hammer}",
      journal = {\pasp},
         year = 2013,
        month = mar,
       volume = {125},
       number = {925},
        pages = {306},
          doi = {10.1086/670067},
archivePrefix = {arXiv},
       eprint = {1202.3665},
 primaryClass = {astro-ph.IM},
       adsurl = {https://ui.adsabs.harvard.edu/abs/2013PASP..125..306F}
}

@ARTICLE{sariesin01,
       author = {{Sari}, Re'em and {Esin}, Ann A.},
        title = "{On the Synchrotron Self-Compton Emission from Relativistic Shocks and Its Implications for Gamma-Ray Burst Afterglows}",
      journal = {\apj},
         year = 2001,
        month = feb,
       volume = {548},
       number = {2},
        pages = {787-799},
          doi = {10.1086/319003},
archivePrefix = {arXiv},
       eprint = {astro-ph/0005253},
 primaryClass = {astro-ph},
       adsurl = {https://ui.adsabs.harvard.edu/abs/2001ApJ...548..787S}
}

@ARTICLE{rhoads99,
   author = {{Rhoads}, J.~E.},
    title = "{The Dynamics and Light Curves of Beamed Gamma-Ray Burst Afterglows}",
  journal = {\apj},
   eprint = {astro-ph/9903399},
     year = 1999,
    month = nov,
   volume = 525,
    pages = {737-749},
      doi = {10.1086/307907},
   adsurl = {http://adsabs.harvard.edu/abs/1999ApJ...525..737R}
}

@ARTICLE{mrees99,
   author = {{M{\'e}sz{\'a}ros}, P. and {Rees}, M.~J.},
    title = "{GRB 990123: reverse and internal shock flashes and late afterglow behaviour}",
  journal = {\mnras},
   eprint = {astro-ph/9902367},
     year = 1999,
    month = jul,
   volume = 306,
    pages = {L39-L43},
      doi = {10.1046/j.1365-8711.1999.02800.x},
   adsurl = {http://adsabs.harvard.edu/abs/1999MNRAS.306L..39M}
}

@ARTICLE{pei92,
       author = {{Pei}, Yichuan C.},
        title = "{Interstellar Dust from the Milky Way to the Magellanic Clouds}",
      journal = {\apj},
         year = 1992,
        month = aug,
       volume = {395},
        pages = {130},
          doi = {10.1086/171637},
       adsurl = {https://ui.adsabs.harvard.edu/abs/1992ApJ...395..130P}
}

@ARTICLE{ryan20,
       author = {{Ryan}, Geoffrey and {van Eerten}, Hendrik and {Piro}, Luigi and {Troja}, Eleonora},
        title = "{Gamma-Ray Burst Afterglows in the Multimessenger Era: Numerical Models and Closure Relations}",
      journal = {\apj},
         year = 2020,
        month = jun,
       volume = {896},
       number = {2},
          eid = {166},
        pages = {166},
          doi = {10.3847/1538-4357/ab93cf},
archivePrefix = {arXiv},
       eprint = {1909.11691},
 primaryClass = {astro-ph.HE},
       adsurl = {https://ui.adsabs.harvard.edu/abs/2020ApJ...896..166R}
}

@ARTICLE{zhang09,
       author = {{Zhang}, Weiqun and {MacFadyen}, Andrew},
        title = "{The Dynamics and Afterglow Radiation of Gamma-Ray Bursts. I. Constant Density Medium}",
      journal = {\apj},
         year = 2009,
        month = jun,
       volume = {698},
       number = {2},
        pages = {1261-1272},
          doi = {10.1088/0004-637X/698/2/1261},
archivePrefix = {arXiv},
       eprint = {0902.2396},
 primaryClass = {astro-ph.HE},
       adsurl = {https://ui.adsabs.harvard.edu/abs/2009ApJ...698.1261Z}
}

@ARTICLE{wang18,
       author = {{Wang}, Xiang-Gao and {Zhang}, Bing and {Liang}, En-Wei and {Lu}, Rui-Jing and {Lin}, Da-Bin and {Li}, Jing and {Li}, Long},
        title = "{Gamma-Ray Burst Jet Breaks Revisited}",
      journal = {\apj},
         year = 2018,
        month = jun,
       volume = {859},
       number = {2},
          eid = {160},
        pages = {160},
          doi = {10.3847/1538-4357/aabc13},
archivePrefix = {arXiv},
       eprint = {1804.02113},
 primaryClass = {astro-ph.HE},
       adsurl = {https://ui.adsabs.harvard.edu/abs/2018ApJ...859..160W}
}

@ARTICLE{panai07,
       author = {{Panaitescu}, A.},
        title = "{Jet breaks in the X-ray light-curves of Swift gamma-ray burst afterglows}",
      journal = {\mnras},
         year = 2007,
        month = sep,
       volume = {380},
       number = {1},
        pages = {374-380},
          doi = {10.1111/j.1365-2966.2007.12084.x},
archivePrefix = {arXiv},
       eprint = {0705.1015},
 primaryClass = {astro-ph},
       adsurl = {https://ui.adsabs.harvard.edu/abs/2007MNRAS.380..374P}
}

@ARTICLE{van12,
       author = {{van Eerten}, Hendrik and {van der Horst}, Alexander and {MacFadyen}, Andrew},
        title = "{Gamma-Ray Burst Afterglow Broadband Fitting Based Directly on Hydrodynamics Simulations}",
      journal = {\apj},
         year = 2012,
        month = apr,
       volume = {749},
       number = {1},
          eid = {44},
        pages = {44},
          doi = {10.1088/0004-637X/749/1/44},
archivePrefix = {arXiv},
       eprint = {1110.5089},
 primaryClass = {astro-ph.HE},
       adsurl = {https://ui.adsabs.harvard.edu/abs/2012ApJ...749...44V}
}

@ARTICLE{kangas20,
       author = {{Kangas}, Tuomas and {Fruchter}, Andrew S. and {Cenko}, S. Bradley and {Corsi}, Alessandra and {de Ugarte Postigo}, Antonio and {Pe'er}, Asaf and {Vogel}, Stuart N. and {Cucchiara}, Antonino and {Gompertz}, Benjamin and {Graham}, John and {Levan}, Andrew and {Misra}, Kuntal and {Perley}, Daniel A. and {Racusin}, Judith and {Tanvir}, Nial},
        title = "{The Late-time Afterglow Evolution of Long Gamma-Ray Bursts GRB 160625B and GRB 160509A}",
      journal = {\apj},
         year = 2020,
        month = may,
       volume = {894},
       number = {1},
          eid = {43},
        pages = {43},
          doi = {10.3847/1538-4357/ab8799},
archivePrefix = {arXiv},
       eprint = {1906.03493},
 primaryClass = {astro-ph.HE},
       adsurl = {https://ui.adsabs.harvard.edu/abs/2020ApJ...894...43K}
}

@ARTICLE{cunningham20,
       author = {{Cunningham}, Virginia and {Cenko}, S. Bradley and {Ryan}, Geoffrey and {Vogel}, Stuart N. and {Corsi}, Alessandra and {Cucchiara}, Antonino and {Fruchter}, Andrew S. and {Horesh}, Assaf and {Kangas}, Tuomas and {Kocevski}, Daniel and {Perley}, Daniel A. and {Racusin}, Judith},
        title = "{GRB 160625B: Evidence for a Gaussian-shaped Jet}",
      journal = {\apj},
         year = 2020,
        month = dec,
       volume = {904},
       number = {2},
          eid = {166},
        pages = {166},
          doi = {10.3847/1538-4357/abc2cd},
archivePrefix = {arXiv},
       eprint = {2009.00579},
 primaryClass = {astro-ph.HE},
       adsurl = {https://ui.adsabs.harvard.edu/abs/2020ApJ...904..166C}
}

@ARTICLE{debarros11,
       author = {{de Barros}, G. and {Amati}, L. and {Bernardini}, M.~G. and {Bianco}, C.~L. and {Caito}, L. and {Izzo}, L. and {Patricelli}, B. and {Ruffini}, R.},
        title = "{On the nature of GRB 050509b: a disguised short GRB}",
      journal = {\aap},
         year = 2011,
        month = may,
       volume = {529},
          eid = {A130},
        pages = {A130},
          doi = {10.1051/0004-6361/201116659},
archivePrefix = {arXiv},
       eprint = {1101.5612},
 primaryClass = {astro-ph.HE},
       adsurl = {https://ui.adsabs.harvard.edu/abs/2011A&A...529A.130D}
}

@ARTICLE{sari99,
       author = {{Sari}, Re'em and {Piran}, Tsvi and {Halpern}, J.~P.},
        title = "{Jets in Gamma-Ray Bursts}",
      journal = {\apjl},
         year = 1999,
        month = jul,
       volume = {519},
       number = {1},
        pages = {L17-L20},
          doi = {10.1086/312109},
archivePrefix = {arXiv},
       eprint = {astro-ph/9903339},
 primaryClass = {astro-ph},
       adsurl = {https://ui.adsabs.harvard.edu/abs/1999ApJ...519L..17S}
}

@ARTICLE{piran04,
       author = {{Piran}, Tsvi},
        title = "{The physics of gamma-ray bursts}",
      journal = {Reviews of Modern Physics},
         year = 2004,
        month = oct,
       volume = {76},
       number = {4},
        pages = {1143-1210},
          doi = {10.1103/RevModPhys.76.1143},
archivePrefix = {arXiv},
       eprint = {astro-ph/0405503},
 primaryClass = {astro-ph},
       adsurl = {https://ui.adsabs.harvard.edu/abs/2004RvMP...76.1143P}
}

@ARTICLE{granot99,
       author = {{Granot}, Jonathan and {Piran}, Tsvi and {Sari}, Re'em},
        title = "{Images and Spectra from the Interior of a Relativistic Fireball}",
      journal = {\apj},
         year = 1999,
        month = mar,
       volume = {513},
       number = {2},
        pages = {679-689},
          doi = {10.1086/306884},
archivePrefix = {arXiv},
       eprint = {astro-ph/9806192},
 primaryClass = {astro-ph},
       adsurl = {https://ui.adsabs.harvard.edu/abs/1999ApJ...513..679G}
}

@ARTICLE{granot01,
       author = {{Granot}, Jonathan and {Loeb}, Abraham},
        title = "{Chromatic Signatures in the Microlensing of Gamma-Ray Burst Afterglows}",
      journal = {\apjl},
         year = 2001,
        month = apr,
       volume = {551},
       number = {1},
        pages = {L63-L66},
          doi = {10.1086/319843},
archivePrefix = {arXiv},
       eprint = {astro-ph/0101234},
 primaryClass = {astro-ph},
       adsurl = {https://ui.adsabs.harvard.edu/abs/2001ApJ...551L..63G}
}

@ARTICLE{vaneerten11,
       author = {{van Eerten}, Hendrik J. and {MacFadyen}, Andrew I.},
        title = "{Synthetic Off-axis Light Curves for Low-energy Gamma-Ray Bursts}",
      journal = {\apjl},
         year = 2011,
        month = jun,
       volume = {733},
       number = {2},
          eid = {L37},
        pages = {L37},
          doi = {10.1088/2041-8205/733/2/L37},
archivePrefix = {arXiv},
       eprint = {1102.4571},
 primaryClass = {astro-ph.HE},
       adsurl = {https://ui.adsabs.harvard.edu/abs/2011ApJ...733L..37V}
}

@ARTICLE{gompertz18,
       author = {{Gompertz}, B.~P. and {Fruchter}, A.~S. and {Pe'er}, A.},
        title = "{The Environments of the Most Energetic Gamma-Ray Bursts}",
      journal = {\apj},
         year = 2018,
        month = oct,
       volume = {866},
       number = {2},
          eid = {162},
        pages = {162},
          doi = {10.3847/1538-4357/aadba8},
archivePrefix = {arXiv},
       eprint = {1802.07730},
 primaryClass = {astro-ph.HE},
       adsurl = {https://ui.adsabs.harvard.edu/abs/2018ApJ...866..162G}
}

@ARTICLE{granot00,
       author = {{Granot}, Jonathan and {Piran}, Tsvi and {Sari}, Re'em},
        title = "{The Synchrotron Spectrum of Fast Cooling Electrons Revisited}",
      journal = {\apjl},
         year = 2000,
        month = may,
       volume = {534},
       number = {2},
        pages = {L163-L166},
          doi = {10.1086/312661},
archivePrefix = {arXiv},
       eprint = {astro-ph/0001160},
 primaryClass = {astro-ph},
       adsurl = {https://ui.adsabs.harvard.edu/abs/2000ApJ...534L.163G}
}



\end{document}